\documentclass[journal=jpccck,articletitle=true,manuscript=article]{achemso}
\usepackage{multirow}
\usepackage{bm}
\usepackage{epsfig}
\usepackage{graphicx}
\usepackage{amsmath}
\usepackage{amssymb}
\usepackage{gensymb}
\usepackage[usenames]{color}
\usepackage{epstopdf}
\usepackage{float}
\usepackage{caption}
\usepackage{xcolor}
\usepackage{soul}
\usepackage{subcaption}
\usepackage{ulem}
\newcommand\colorsout[1]{\bgroup \markoverwith{\textcolor{#1}{\rule[0.5ex]{2pt}{0.4pt}}}\ULon}

\author{R. E. Pico} \affiliation{Instituto de F\'{\i}sica de Rosario, Consejo Nacional de Investigaciones Cient\'{\i}ficas y T\'ecnicas (CONICET) and Universidad Nacional de Rosario, Bv. 27 de Febrero 210 Bis (2000) Rosario, Argentina} \email{pico@ifir-conicet.gov.ar}
\author{A. F. Rebola}
\affiliation{Instituto de F\'{\i}sica de Rosario, Consejo Nacional de Investigaciones Cient\'{\i}ficas y T\'ecnicas (CONICET) and Universidad Nacional de Rosario, Bv. 27 de Febrero 210 Bis (2000) Rosario, Argentina}

\author{J. Lasave}
\affiliation{Instituto de F\'{\i}sica de Rosario, Consejo Nacional de Investigaciones Cient\'{\i}ficas y T\'ecnicas (CONICET) and Universidad Nacional de Rosario, Bv. 27 de Febrero 210 Bis (2000) Rosario, Argentina}

\author{P. Abufager}
\affiliation{Instituto de F\'{\i}sica de Rosario, Consejo Nacional de Investigaciones Cient\'{\i}ficas y T\'ecnicas (CONICET) and Universidad Nacional de Rosario, Bv. 27 de Febrero 210 Bis (2000) Rosario, Argentina}

\author{I. J. Hamad}
\affiliation{Instituto de F\'{\i}sica de Rosario, Consejo Nacional de Investigaciones Cient\'{\i}ficas y T\'ecnicas (CONICET) and Universidad Nacional de Rosario, Bv. 27 de Febrero 210 Bis (2000) Rosario, Argentina}
\email{hamad@ifir-conicet.gov.ar}

\date{\today}

\title[]{\bf Modelling the magnetic properties of 1D arrays of FePc molecules}

\begin{document}

\begin{tocentry}
\begin{center}
\includegraphics*[scale=0.295]{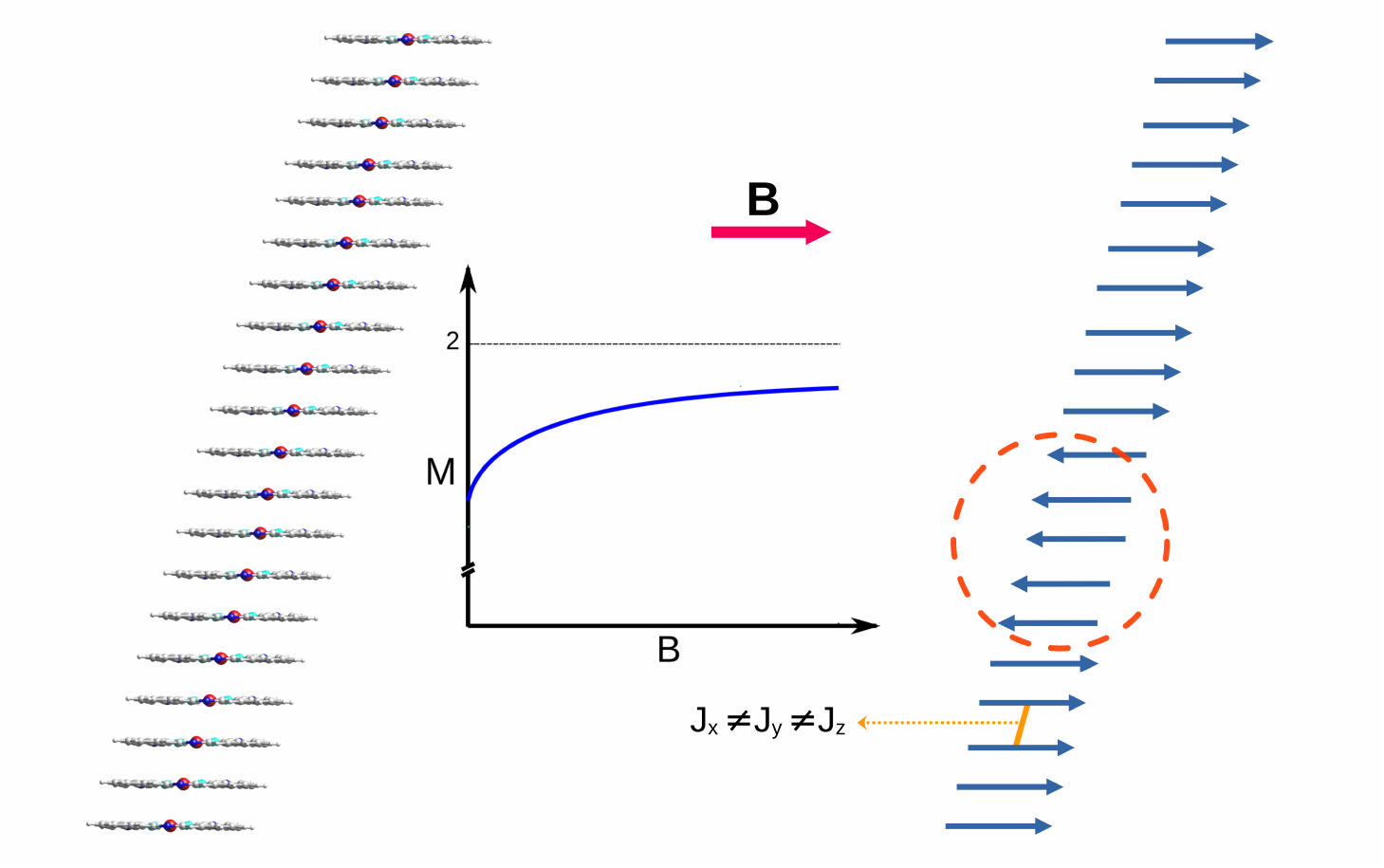}
\end{center}
\end{tocentry}

\begin{abstract}
We investigate the magnetic properties of Fe Phthalocyanines (FePc) that are experimentally arranged in quasi one-dimensional chains when they  are grown in thin films or powders. By means of DFT calculations we reproduce the structural parameters found in experiments, and then we build a generalized Heisenberg magnetic model with single ion anisotropy, and calculate its parameters. The results show a anisotropic exchange interaction $J$ between FePc molecules, and an easy plane single ion anisotropy $D$. By means of Monte Carlo simulations, with this model, we found an explanation to the non-saturation of the magnetization found at high fields, which we interpret is due to the anisotropic exchange interaction $J$. Finally, we also investigate the presence of magnetic solitons versus temperature and magnetic field. This results provide additional evidence that FePc is a soliton bearing molecular compound, with solitons easily excited mainly in the molecular $xy$ plane.

\end{abstract}

\newpage

\section{Introduction} \label{Introduction}

Over the past decade, significant progress has been made in the field of organic materials 
due to their potential applications as nanoscale spintronics \cite{Guo2019}, 
organic photovoltaics \cite{Inganas2018, Grant2019}, transistors \cite{Fleet2017,Klauk2010,Zhou2021}, 
light-emitting diodes \cite{Kalyani2012,Plint2016}, and electrochemical devices \cite{Rivnay2018,Basova2020}. Organic materials constructed with one-dimensional building blocks, in particular, have become a subject of great interest because they serve as excellent model systems for studying and applying low dimensional magnetic phenomena. It is important, then, to understand the underlying principles of their magnetic properties. Among the building blocks for organic materials, metal phthalocyanines (MPcs) are one of the most extensively used \cite{Cranston2021}.

\begin{figure*}[ht]
 \includegraphics[width=15.1cm]{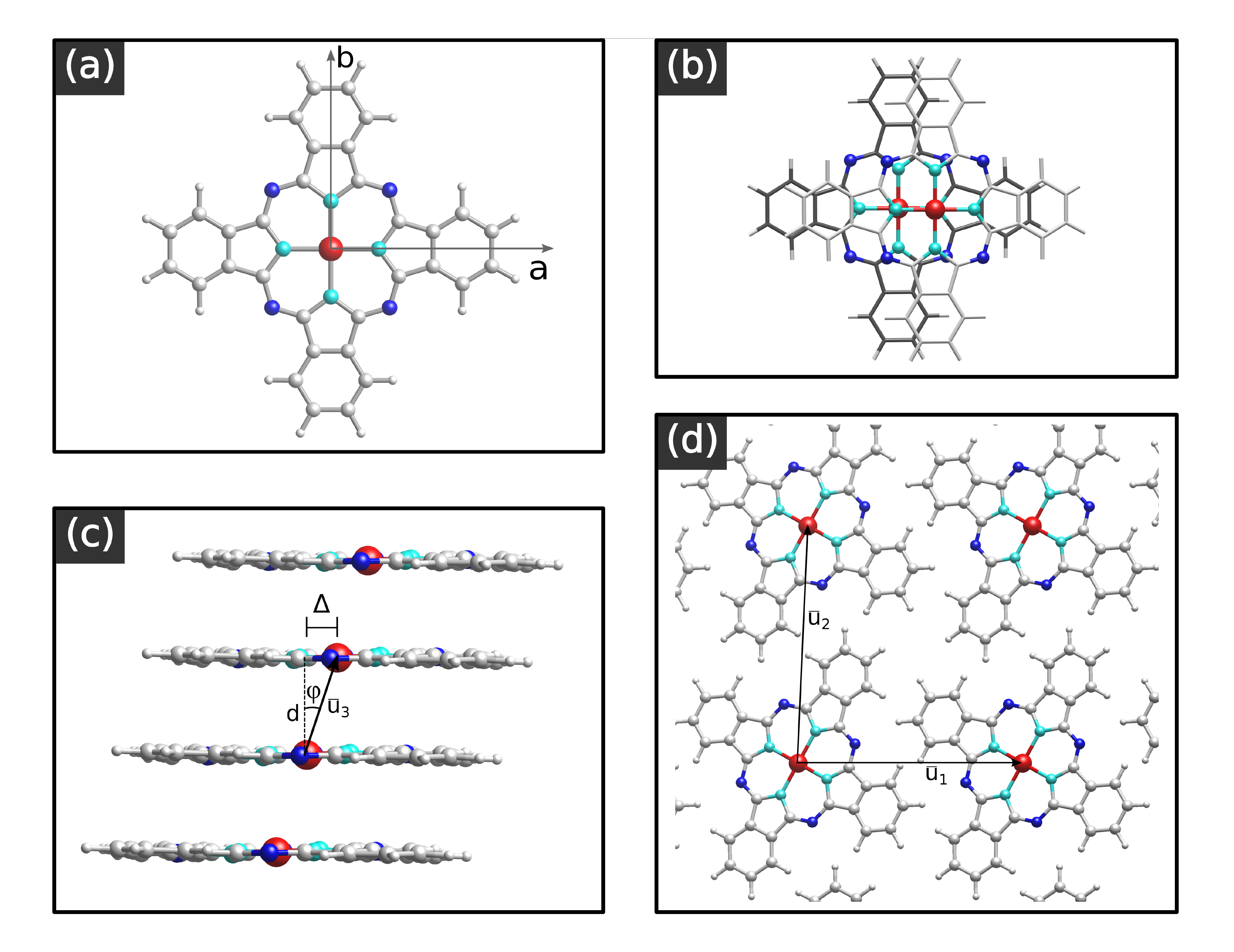}
  \caption{  (a) FePc molecule view from top. The four Nitrogen closer and farthest to the  Fe atom  are  the pyrrole (N$_{py}$) and the bridging (N$_{bridge}$) nitrogen atoms, respectively. Fe, N$_{py}$, N$_{bridge}$, C and H are indicated by red, cyan, blue, gray and white balls, respectively. (b) and (c) Stacking geometry of FePc-film.  $d$ is the distance between the molecular planes and $\varphi$ is the angle between the stacking axis $\bar{u}_3$  and a vector perpendicular to the molecular plane. (d) The cell parameters for a supercell containing 1 molecule are $|\overline{u}_1|=$13.24 \AA, $|\overline{u}_2|=$13.43 \AA, $|\overline{u}_3|=$3.60 \AA,  $(\overline{u}_1,\overline{u}_2)$=85$^{\circ}$, $(\overline{u}_1,\overline{u}_3)$=63.6$^{\circ}$, $(\overline{u}_2,\overline{u}_3)=$74.1$^{\circ}$. It is important to take into account that for the magnetic model,  we employed a cell size that is doubled in the $\overline{u}_3$ direction and contains two molecules per cell. }
 \label{fig1}
 \end{figure*}

The MPc molecule is composed of a conjugated planar macrocycle, MN$_8$C$_3$H$_{16}$, with M being a central metal atom that is 4-coordinated to the N isoindole or pyrrole N atoms, N$_{py}$. The macrocycle is made up of four isoindole units linked by a ring of four N atoms (N$_{bridge}$), as shown in Fig.\ref{fig1}(a). Depending on the growth conditions, MPc crystallizes into a range of polymorphs, that have been observed in powders, thin films (TF), and nanowires systems \cite{Cranston2021}. The building blocks of these crystalline forms are one dimensional (1D) columns, composed of MPc molecules, with the metal atom M at the center serving as the column axis, or the stack axis (see Fig.\ref{fig1}(d)). The stacking columnar structures differ in the tilting angles ($\varphi$)  between the stacking axis and the molecular plane, and the interplanar distance ($d$) between two neighboring molecular planes along the column, resulting in different Pc crystal structures (see Fig.\ref{fig1}(c))
\cite{Bartolome2014,Cranston2021,Hoshiro2003,Wu2019,Yim2003,Wang2010,Bartolome2015}. Modifying the magnetic center of MPc introduces an additional  degree of freedom to the properties of its polymorphs, since it impacts the magnetic properties of the molecule and hence of the crystals formed by them.

Two different arrangements have been proposed to describe the stacking geometry of MPc molecules. The first one is referred to as the herringbone structure, where the molecules stack with opposite angles between adjacent columns \cite{Ashida1966}. In the second one, the MPc molecules stack in the same orientation in each column, forming the so-called brickstack structure \cite{Hoshiro2003}.  

Particularly, iron phthalocyanines (FePc) are a class of MPc molecules that have garnered significant attention. In bulk arrangements, FePc exhibits two polymorphs, $\alpha$ and $\beta$, with stacking angles of 26.5$^\circ$ and 44.8$^\circ$, respectively. Previous studies of the $\beta$ phase have shown that it behaves paramagnetically \cite{Evangelisti2002}. However, in the $\alpha$ phase found in powder systems, FePc shows ferromagnetic correlations below approximately 10 K \cite{Evangelisti2002}. The evidence in powders points to an arrangement of the herringbone type \cite{Filoti2006}. On the other hand, in thin films (TF), FePc molecules adopt a brickstack structure when grown on a variety of substrates \cite{Bartolome2015,Bartolome2014,Wu2019,Vargas2020}. The polymorphism found in TF has been called $\alpha_{+}$ \cite{Hoshiro2003} (but it has also been directly called $\alpha$ \cite{Wu2019} or TF \cite{Bartolome2015}), since, besides the brickstack arrangement, it is different from the bulk $\alpha$ phase from the way each molecule is shifted with respect to the one in the upper/lower molecular plane (see section Stacking Geometry below for more detail). In this work we will refer to this polymorphism as  $\alpha_{+}$. FePc thin films have been reported to become ferromagnetically correlated at temperatures below 20 K \cite{Wu2019} and exhibit a ferromagnetic transition at $4.5$ K \cite{Gredig2012}, very similar to the transition temperature in powders at $5$ K \cite{Bartolome2014}.

In order to explain the magnetic phenomena observed in experiments, both in powders and in TF, different  model Hamiltonians have been proposed. Moreover, even the same parameters have very different proposed values in different works (see section Magnetic Model below for more details). On the experimental side, magnetization measurements made in powders and also also in TF show that at relatively high magnetic fields, the magnetization does not saturate \cite{Evangelisti2002,Wu2019}. In the case of powders, this non saturation has been attributed to the herringbone arrangement, due to the canting of the adjacent columns of FePc molecules \cite{Evangelisti2002}. However, in thin films the arrangement is of the brickstack type\cite{Filoti2006}, so that other mechanism must be also contributing to the non saturation of the magnetization as well. Finally, in a recent work, experimental evidence of magnetic solitons has been found \cite{Wu2019}. Since FePc has gained attention partly because of their electrical and magnetic properties, it is important to study how magnetic solitons are developed in this system. In order to shed light on these issues, a theoretical study beginning from first principles calculations ending in a magnetic model was carried out, to help to understand this and related systems further. \\  

In this work we theoretically study the $\alpha_{+}$ phase of FePc chains, by means of  density functional theory (DFT) and classical Monte Carlo. With DFT calculations, we first obtain structural parameters, comparing with available experimental data, and then proceed to calculate the parameters of a generalized Heisenberg model with single ion anisotropy. We obtain a  anisotropic Heisenberg model, with an easy plane single ion magnetic anisotropy $D$ and negligible $E$. Afterwards, we study this model in an array of weakly coupled FePc chains, by means of classical Monte Carlo simulations. We were able to explain the lack of saturation observed in magnetization experiments. Finally, we investigate the magnetic excitations of this system, focusing on exploring the existence of solitons in the phase diagram. \\

\section{Methods}
 \subsection{\label{sec:level1}Computational Details }
 \label{DFT_calculations}

DFT calculations were performed with the VASP code 
\cite{Kresse1993a,Kresse1993b,Kresse1996a,Kresse1996b,Kresse1999,Hafner2008} within 
the slab-supercell approach and using the projector augmented-wave (PAW) method \cite{Kresse1999}.Wave functions were expanded using a plane wave basis set with an energy cut-off of 550 eV. Different schemes were tested in order to improve the description of vdW interaction: the PBE+D3 \cite{Grimme2010} approach and the optB86b-vdW, optB88-vdW, optPBE-vdW, vdW-DF2 non-local correlation functionals \cite{Klimes2011}. From test calculations  for the bulk-like FePc structure, we conclude that the optb86b scheme is the one that gives the best interplanar distance with respect to the experimental value, therefore, it is  used for the structural determination of bulk structures. However, when calculating the parameters of the magnetic model, in particular the anisotropy parameters $D$ and $E$, which require spin-orbit interaction turned on, opt86b is not suitable, and the PBE+D3  scheme is used instead (see section magnetic model). Exchange coupling parameters $J_x$, $J_y$ and $J_z$ are also calculated within this PBE+D3 scheme. For testing purposes, in an isotropic Heisenberg model the exchange coupling $J$ has no appreciable difference when calculated with \textit{PBE$+$D3} or with optb86b.

The Hubbard U correction is considered with the DFT+U approximation \cite{Dudarev1998} to deal at DFT level with the Fe d-electrons. $U_{eff}=U-J=3 eV$ was chosen, since this value has also been used in previous studies in FePc molecules\cite{Roberto2012,Wang2018}. 
 
The FePc $\alpha_{+}$ polymorphism was studied. For testing purposes, results for the gas phase of FePc are also presented here. Gas phase calculations were carried out at $\Gamma$ point. For the $\alpha_{+}$ phase, the \textit{k-point} grid was 2x2x15. Calculation without and with spin-orbit coupling (SOC) have been performed. Calculations with SOC were carried out at fixed geometry. In all cases, $k$ point convergences were achieved. All geometry optimizations were carried  out until the forces on  every mobile atom were smaller than 0.02 eV/\AA, and all calculations were spin polarized.

\subsection{Gas Phase calculations}
\label{GS}

Despite the numerous theoretical and experimental studies there is still no definite consensus about the ground state of the molecule. However, there is a broad consensus that the FePc molecule in gas phase has spin $S = 1$. DFT studies have predicted $^3A_{2g}$ [$(d_{z^2}^{\uparrow \downarrow}) \ (d_{xz,yz}^{\uparrow \uparrow}) \ (d_{xy}^{\uparrow \downarrow})$]  \cite{10Marom2009} , $^3E_g(a)$ [$(d_{z^2}^{\uparrow}) \ (d_{xz,yz}^{\uparrow \downarrow \uparrow}) \ (d_{xy}^{\uparrow \downarrow})$] \cite{40Bialek2003,41Kuzmin2009} and $^3B_{2g}$ [$(d_{z^2}^{\uparrow}) \ (d_{xz,yz}^{\uparrow \downarrow \uparrow \downarrow}) \ (d_{xy}^{\uparrow})$] \cite{Brena2011} as possible ground states and the prediction actually depends on the choice of the exchange-correlation (XC) functionals \cite{Marom2009}.  The same conclusion has been established by a recent work based on a diffusion Monte Carlo study where they obtain mainly $^3A_{2g}$, but also $^3B_{2g}$ and even $^3E_g$ depending on XC \cite{Ichibha2017}. A recent theoretical work employing numerous and sofisticated methodologies has concluded that the ground state of the molecule is $^3A_{2g}$ \citep{Phung2023}.

Within our calculation, we obtained $^3E_g(a)$ as the ground state (GS), a $^3A_{2g}$ state at 36 meV and $^3B_{2g}$ at 279 meV higher energy in line with the results obtained by Ichiba et al. \cite{Ichibha2017} 
for DFT+U calculations with U$=$4eV, and also with Ref \cite{Nakamura2012}. 
However, as mentioned earlier, in the isolated molecule, numerous configurations appear to be closely competitive. For instance, in our case, the energy difference between $^3E_g(a)$ and $^3A_{2g}$ is merely 36 meV. Sophisticated methods like those of Refs. \citep{Ichibha2017, Phung2023} need to be employed to obtain a definite answer to the GS of the isolated molecule. Our focus will be on FePc chains on TF, where the situation seems to be different (see below). Finally, for all electronic configurations, the spin magnetic moment of the molecule is close to 2$\mu_b$ concentrated in the Fe center. 

\begin{figure}[h]
 \includegraphics[width=8.3cm]{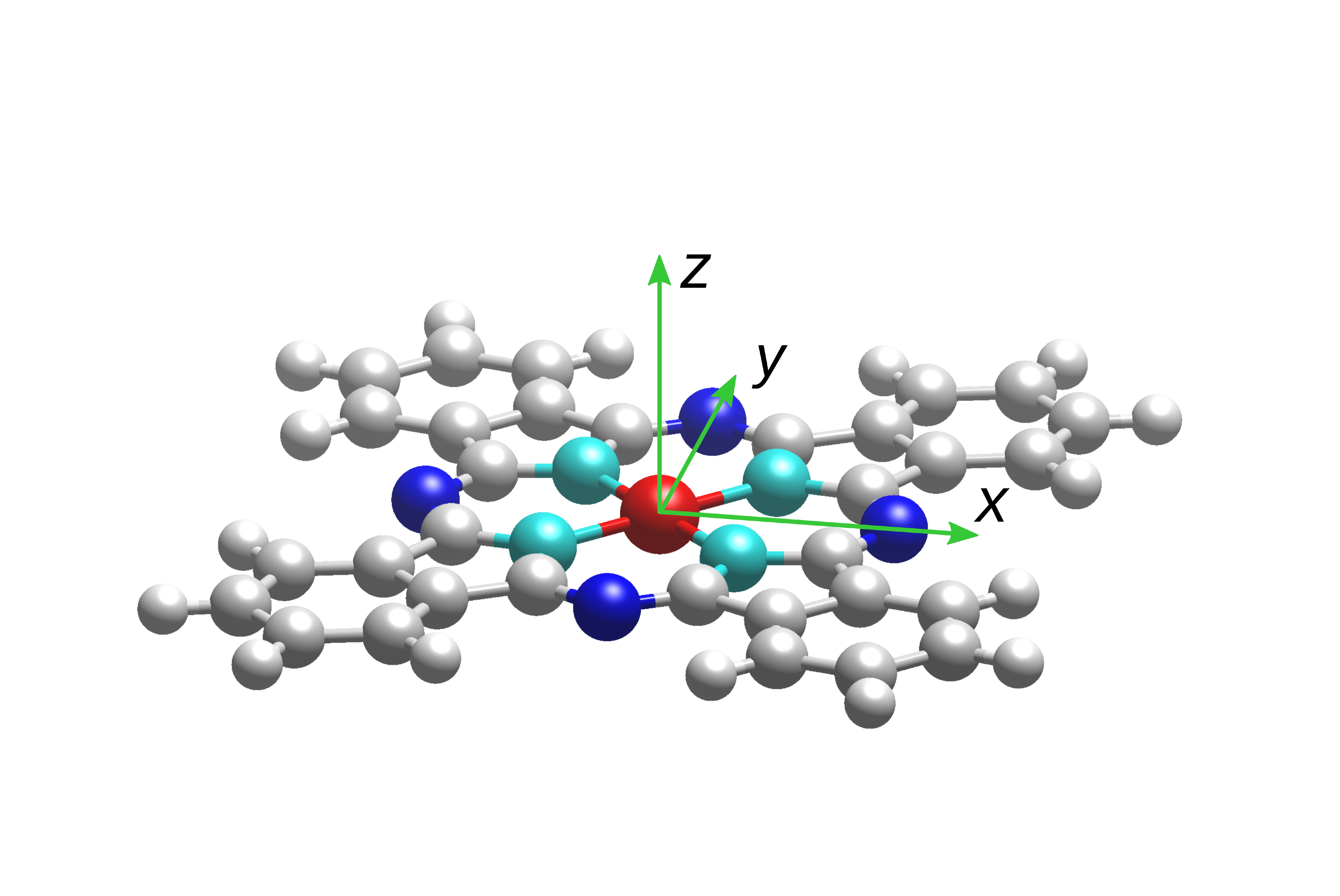}
\caption{Directions were the total energy was computed in DFT calculations including SOC. Perpendicular to the molecular plane (001) and two in plane orthogonal directions (100 and 010), respectively.}
 
\label{vectors}
\end{figure}

\subsection{$\alpha_+$ phase. Stacking Geometry}
\label{geometry}

In FePc TF grown on different substrates, molecules stack layer by layer almost parallel (with a tilting angle of $7$ or $9$ degrees) to the substrate surface  \cite{Bartolome2015, Wu2019, Vargas2020}. In this "flat-lying" configuration, two adjacent molecular layers are shifted a distance $\Delta$ (see Fig. \ref{fig1}(c)) apart from one another along the Fe-N$_{py}$ bond direction indicated by the $a$, or alternatively the $b$, axis in Fig.\ref{fig1}(a). As a result, the Fe atom in a molecule is close to the center of a N$_{py}$ atom in the molecule beneath it \cite{Bartolome2015, Wu2019} (see Fig.\ref{fig1}(b)). In this stacking geometry Fe atoms form parallel unidimensional chains that deviate from the normal to the molecular plane by an angle $\varphi$ (see. Fig.\ref{fig1}(c)). In the case of the herringbone arrangement found in powders, apart from the two-sublattice canted structure the other difference is that the shift between FePc molecules occurs along the Fe-N$_{bridge}$ bond \cite{Bartolome2015}. In the case of a FePc TF grown on Au(111)  \cite{Bartolome2015}, a structural refinement based on XLPA simulations gives  $\Delta$=1.48 \AA, which is very close to the estimated value from STM images, $\Delta$=(1.0 $\pm$ 0.3) \AA. The distance between adjacent molecular planes is $d_z$ = 3.25 \AA, and molecules within the molecular plane form a square lattice with a lattice parameter of 13.0 $\pm$ 0.2 \AA \cite{Bartolome2015} (see Fig.\ref{fig1}(d)).

The estimated experimental values for FePc TF grown on PTCDA/Si(100) \cite{Wu2019} are $d_z $ = 3.42 \AA \ and a distance  $d_{Fe-Fe} =$ 3.77 \AA \ between the nearest Fe atoms within a chain. In the case of FePc TF grown Au-coated sapphire, $d_z$= 3 \AA  and the lateral separation between chains is about 13 \AA \cite{Vargas2020}. Similar values, with slightly varying, thickness-dependent, stacking angles, have been found in a recent study of FePc films in different substrates \cite{Xia2023}.  The similarities in structural data reported for FePc-TF in the "flat-lying" configuration suggest that once layer-by-layer growth is initiated due to the templating effect of the substrate \cite{Yim2003}, intermolecular interactions seem to drive the film structure. This leads to performing DFT structural calculations without including the substrate.

We have performed DFT calculations for bulk-like structures with one molecule per unit cell, simultaneously optimizing the unit cell and the molecular degrees of freedom, with the aim of reproducing to the most FePc-TF structure. For the obtained lowest energy structure, the shift of two adjacent layers runs  along  Fe-N$_{py}$ bond axis\footnote{The deviation is less than 2$^{\circ}$}, with a distance of $\Delta=1.61$ \AA. The distance between molecular planes,  the stacking angle of the Fe chain and Fe-Fe distance within a chain are, respectively, $d_z=3.22$ \AA,  $\varphi=26.6^{\circ}$ and $d_{Fe-Fe}=3.60$ $\AA$. Furthermore, the dimensions of the unit cell vectors u$_1$ and u$_2$ and the angle among them  are 13.2 \AA, 13.4 \AA \ and 85$^{\circ}$, respectively. Because the molecular plane is close to the defined plane by u$_1$ and u$_2$  \footnote{The angular deviation is $\sim$4$^{\circ}$} their dimensions can be compared to the lateral distance between chains experimentally measured. Overall, all parameters agree well with the previously described experimental data, lending support to the DFT calculations for the $\alpha_+$ phase. For this stacking geometry, the calculations show that each molecule is in a  $E_g(a)$ configuration 
($(d_{z^2}^{\uparrow}) \ (d_{xz,yz}^{\uparrow \downarrow \uparrow}) \ (d_{xy}^{\uparrow \downarrow})$). This result coincides with previous constrained DFT calculations \cite{Nakamura2012} and with experimental evidence \cite{Bartolome2010, Filoti2006}, altough other works predict a ground state in TF that is a superposition of $^3A_{2g}$,$^3B_{2g}$ and $^3E_{g}$ \cite{Natoli2018}.   

Finally, the calculated spin magnetic moment in the Fe atom is  $m_s (Fe)=2.021\mu_b$, corresponding to $S=1$, the same than in the gas phase.  

\section{Results and discussion}
\subsection{Magnetic model}
\label{magnetic model}

On the basis of experimental evidence, several magnetic models have been proposed for the bulk phase $\alpha$ and for the TF $\alpha_+$ of FePc. There is consensus that the molecule has $S=1$, and also that there is a positive single ion magnetic anisotropy $D$, aside from the ferromagnetic exchange $J$. Working  with powders of FePc molecules, Evangelisti {\it{et. al}}\cite{Evangelisti2002} observed that the magnetization did not saturate even at high magnetic fields, and this was interpreted in terms of the above mentioned herringbone arrangement, a canted structure between two different magnetic sublattices originally proposed for $MnPc$. An effective Ising $S=1/2$  model was proposed. This Ising model was later backed up in Ref. \cite{Filoti2006}. However, it was not considered as appropriate to account for the observed behaviour in TF in Ref. \cite{Wu2019}, and an isotropic Heisenberg model with single ion anisotropy was proposed instead. As for the mechanism for the magnetic exchange, in Ref. \cite{Evangelisti2002} the direct Fe-Fe interaction was proposed, which renders the interaction ferromagnetic in character. However, the value proposed for the exchange parameter $J/k_B$ or $J_z/k_B$ varies significantly. In a Ising model written as $-2J_z \sum_i S_{z,i} S_{z,i+1}$ for powder samples, it was estimated to have the value of $25.7 K$ in Ref. \cite{Evangelisti2002}  and $76 K$ in Ref. \cite{Filoti2006} . In isotropic Heisenberg models $-J \sum_i S_{i} S_{i+1}$, $J/k_B$ was estimated to be $20 K$ in Ref. \cite{Wu2019} and $13 K$ in Ref. \cite{Xia2023}, these last two working on TF.  On the other hand, while there is a consensus that the magnetic anisotropy $D$ is positive, indicative of an easy plane situation, its value has been estimated from experiments, varying from $D=92 $ K  in Ref. \cite{Barraclough1970} to $D = 53 K$ in Ref. \cite{Wu2019}. 

Spin orbit interaction energy is important in FePc molecules, since they present a high unquenched orbital moment \cite{Bartolome2010, Filoti2006}. In this case, it is well known that an effective spin hamiltonian can be derived, which contains terms of the form $D S_z^2 + E (S_x^2 + S_y^2)$, being the parameters $D$ and $E$ matrix elements proportional to the orbital magnetic moment. The $D_{4h}$ symmetry present in isolated FePc molecules is broken when the chains are formed in the bulk or thin film structure, since the stacking axis is not perpendicular to the molecular plane. This suggests the idea that the Heisenberg exchange term may in principle be anisotropic. An extreme case of this is precisely the Ising model considered in Ref. \cite{Evangelisti2002}. However, most recent works in FePc bulk structures or TF have considered the isotropic Heisenberg model only \cite{Wu2019, Vargas2020, Vargas_2020_2}. On the other hand, the interchain exchange is expected to be relatively small, due to relatively large lateral distances between Fe centers, as confirmed by calculations done in CrPc chains \cite{Wu2013}. These considerations render FePc thin films to be treated, as DFT calculations are concerned, as one dimensional chains of FePc molecules, geometrically arranged as described in section Stacking Geometry, with the possibility of anisotropic magnetic exchange and with other spin-orbit derived interactions present, such as single ion magnetic anisotropy terms. On the other hand, regarding the Dzyaloshinskii-Moriya (DM) anisotropic exchange, it's important to consider that its order of magnitude is given by  $D_{DM} \sim (\Delta(g)/g) J$ \cite{Moriya60}. With the most recent estimation of $g$ in TF, $g=2.16$ \cite{Wu2019} the magnitude can be estimated in $D_{DM} \sim 0.08 J$. Consequently, this interaction has not been considered significant in this work. \\  

With all this considered, the proposed model Hamiltonian for the one dimensional chains of FePc molecules in the $\alpha_+$ polymorphism is: 

\begin{equation}
    \begin{split}
     H&=\sum_i D(S_i^z)^2+\sum_i E(S_{ix}^2-S_{iy}^2)\\
     &-\frac{1}{2}\sum_{<i,j>} J_z(S_i^z \cdot S_j^z)
    -\frac{1}{2}\sum_{<i,j>} J_x(S_i^x \cdot S_j^x)\\
    &-\frac{1}{2}\sum_{<i,j>} J_y(S_i^y \cdot S_j^y)\hspace{2cm}\\
\end{split}
\label{Hamiltoniano}
\end{equation}

\noindent where the sum over i runs over all Fe atoms in the supercell and the notation $<i,j>$ represents first neighbor atoms. The factor $1/2$ is introduced to avoid double counting. The first term in the Hamiltonian describes the uniaxial anisotropy as we choose z as the off-plane direction. The second term describes the in-plane anisotropy. The anisotropic exchange interaction is considered in the last three terms.\\

We now proceed to compute the magnitude of the parameters of the model Hamiltonian with DFT calculations, by means of the widely used method of energy mapping \cite{Guterding2016,Whangbo2017}. To do so, the energy difference between classical configurations, described in Ec. \ref{Energias}, has been computed. 

\begin{equation} 
\begin{split}
    &E_{x}^{FM}=2ES^2-2J_xS^2+E_0\\
    &E_{x}^{AF}=2ES^2+2J_xS^2+E_0\\
    &E_{y}^{FM}=-2ES^2-2J_yS^2+E_0\\
    &E_{y}^{AF}=-2ES^2+2J_yS^2+E_0\\
    &E_{z}^{FM}=2DS^2-2J_zS^2+E_0\\
    &E_{z}^{AF}=2DS^2+2J_zS^2+E_0\\
\end{split}
\label{Energias}
\end{equation}

Such calculations were performed for a chain defined by two molecules per unit cell. For each configuration the directions x, y, z are defined as in Figure \ref{vectors}. In each line of Eq. \ref{Energias}, $E_{direction}^{FM/AFM}$ is defined, where $FM$ means that the magnetic moments of both molecules are parallel and $AFM$, antiparallel. For determining the parameters $J_x$, $J_y$, $J_z$, $D$ and $E$ we used the ground state energy obtained from our DFT calculations for each configuration in equation \ref{Energias}. Then, the parameters expressed as function of the ground states energies were

\begin{equation}
    J_i=\frac{E_i^{AF}-E_i^{FM}}{4S^2}
    \label{equationJ}
\end{equation}
\begin{equation}
    D=\frac{1}{2}\big[E_z^{FM}-\frac{E_x^{FM}+E_y^{FM}}{2}+2(J_z-\frac{J_x+J_y}{2})\big]
    \label{EquationD}
\end{equation}
\begin{equation}
    E=\frac{1}{4}\big[E_x^{FM}-E_y^{FM}+2(J_x-J_y)\big]
    \label{EquationE}
\end{equation}

where the index $i$ runs for $x,y,z$.

The obtained results for the Heisenberg exchange were $J_x=1.77$ meV, $J_y=2.31$ meV and $J_z=2.07$ meV. The precision of the DFT calculations in VASP was more than enough to assure the last digit. While the mean value of the exchange is very similar to the proposed $J$ values in isotropic models, our calculations show that it is anisotropic, with an important amount of exchange anisotropy $J_y/J_x=1.31$. It is important to note that the positive value of the Heisenberg exchange means a ferromagnetic coupling, in line with previous experimental results \cite{Evangelisti2002,Wu2019}. The type and value of the exchange coupling in MPCs is strongly influeced by the stacking angle, hence by the $\Delta$ value\cite{Wu2013}.

The calculated single ion anisotropy is $D=0.65 $ meV, which is equivalent to $\sim 7.5 K$, a value much lower than the estimations from experiments previously mentioned. Nevertheless, it is important to note that the calculated value agrees with earlier theoretical findings in which the preferred in plane magnetization direction was also identified \cite{Nakamura2012}. The easy plane anisotropy obtained is due to the spin-orbit interaction trying to align the spin with the orbital magnetic moment, this being the result of the orbital motions of the Fe electrons. Our results show an orbital magnetic moment greater in the molecular plane, with values of 0.118, 0.140, and 0.039 $\mu_B$ along the $x$, $y$, and $z$ directions, respectively. However, these values are likely underestimated, as suggested by previous research\cite{Bartolome2010}. DFT calculations tend to underestimate the orbital magnetic moment when electron correlation effects are important \cite{Wackerlin2022, Gallardo_2019}, and it is expected that the actual value of $D$ is higher than in our calculations. Finally, the calculations showed than $E$ is very small compared to the other parameters of the model, and consequently it will be disregarded from now on.\\

To this one-dimensional model for FePc chains we add interchain interactions $J_{inter}$ that are expected to be one order of magnitude smaller than the intrachain ones, due to the large interchain Fe-Fe distance compared with the intrachain one \cite{Wu2019,Wu2013}. Although the DFT calculations were made for the particular geometry of the $\alpha+$ phase found in TF, we expect the model to account for the physics of other polymorphisms as well. 
Our results are summarized in table \ref{parametros}. 

  \begin{table}[h]
   \begin{center}
  \begin{tabular}{|c|c|c|c|c|c|}
  \hline
   $\alpha_+$ & $J_x$ & $J_y$ & $J_z$ & D & E   \\
  \hline
  \ (meV) \ & \ 1.77 \ & \ 2.31 \ & \ 2.07 \ & \ 0.65 \ & \ 0.04 \ \\
  \hline
  (K) & \ 20.5 \ & \ 26.8 \ & \ 24.0 \ & \ 7.5 \ & \ 0.5 \ \\
  \hline
  \end{tabular}
  \caption{DFT calculated parameters for the magnetic model of FePc TF $\alpha_+$ phase.}
 \label{parametros}
 \end{center}
 \end{table} 
 
\subsection{Monte Carlo calculations.}
\label{Monte Carlo}

Taking into account the ferromagnetic ground state of weakly coupled FePc chains, with individual molecules having spin $S=1$,  classical Monte Carlo is a suitable method to study its magnetic properties. All the Monte Carlo simulations presented in this work were performed on periodic FePc chains within a $3$x$3$x$1000$ supercell. In such an arrange each spin, representing the Fe magnetic moment, interacts with 2 nearest neighbors within its chain and with 4 nearest neighbors belonging to different chains located along the directions of the x and y axes. The model Hamiltonian is the one described in Eq. \ref{Hamiltoniano}, with the derived parameters: $J_{1x} = 1.76$ meV, $J_{1y} = 2.31$ meV, $J_{1z} = 2.07$ meV (equivalent to $20.5, 27$ and $24$ Kelvin, respectively), magnetic anisotropy $D=0.65$ meV with the addition of interchain interactions of $J_{inter}$, which were shown to be relatively small but not negligible \cite{Wu2019,Wu2013}. The chosen value is $J_{inter}=0.01$ meV, since for this value the transition temperature (i.e. the temperature at which there is a peak in susceptibility or specific heat curves, not shown) is of around 5 Kelvin, a value very similar to that reported in experiments. 

\subsection{Magnetization.}

In magnetization experiments both in powders and thin films, the results showed that even at high fields of several Tesla, the magnetization per molecule did not saturate to the expected value for $S=1$, that is $2 g \mu_B$. In FePc powders, this was interpreted in terms of the herringbone structure (see above), due to the canting of the sublattices \cite{Evangelisti2002}. On the other hand, in thin films, there is evidence that the structure is of the brickstack type \cite{Wu2019}, so that no canted two sublattice structure is present. All the same, the magnetization in thin films does not saturate at high fields. This suggests that another mechanism could be playing a role in this observed behaviour. \\

Concretely in Ref. \cite{Wu2019}, working with TF, it is shown that even at the lowest temperatures reached, of $2 K$: 
\begin{itemize}
    \item For the field in the molecular plane (Fig. 4 d, e, templated films), the magnetization did not saturate for fields as high as 7 Tesla, reaching $1.8 \mu_B$ at this value of the field.
    \item For the field perpendicular to the molecular plane (Fig. 4 a, b, non-templated films), the magnetization is lower than that corresponding to the same value of $B$ for $B$ in the molecular plane ($M <1 \mu_B$ for $B< 6$ Tesla).
\end{itemize}


In the experiments in thin films, grains are formed, so that when the field is in the molecular plane, it can in principle form any angle with the molecular 'y' axis defined above. To simulate this, we have performed Monte Carlo simulations with the magnetic field $B$ at varying angles with the $y$ axis of the molecule, and we have averaged the magnetization, in the direction of the field, for all the directions of $B$ between $0$ and $90$ degrees.  This gives the results showed in table \ref{magnetizacion_B}, all for a $T=2$ Kelvin, the lowest reported in the experiments.  

\begin{table}[h]
   \centering
   \caption{Average magnetization (in Bohr magnetons) for different values of $B$, being the field in the molecular plane always. The Monte Carlo calculations correspond to $T=2 \ K$.}
   \label{magnetizacion_B}
   \small 
   \begin{tabular}{|c|c|c|c|c|c|c|c|}
   \hline
   B (T) & 1 & 2 & 3 & 4 & 5 & 6 & 7 \\
   \hline
   $M_{avg} (\mu_B)$ & 1.30 & 1.39 & 1.47 & 1.54 & 1.61 & 1.66 & 1.73 \\
   \hline
   $M_{avg}/M_{sat}$ & 0.65 & 0.70 & 0.73 & 0.77 & 0.80 & 0.83 & 0.86 \\
   \hline
   \end{tabular}
\end{table}

The results are in very good agreement with those presented in Ref. \cite{Wu2019}. For fields below $3$ Tesla, the magnetization is below $1,5 \ \mu_B$, i.e. $75 \%$ of the saturation vale, and for a field of magnitude $7$ Tesla, we obtain a magnetization of $1,73 \ \mu_B \equiv 86 \% M_{sat}$, close to the $1,8 \ \mu_B$ reported in experiments that represented $83 \%$ of the saturation value. Thus, in the frame of our model, the magnetization does not saturate even at fields of 7 Tesla, due to the anisotropy of the exchange $J$. \footnote{Within our model, saturation is indeed reached at much larger fields, and the non-saturation of the magnetization in FePc powders at $B=20$ Tesla reported in Ref. \cite{Evangelisti2002} has to be explained adding the canting of the FePc sublattices found in the herringbone phase.}    \\

With our model, the non saturation of magnetization within the molecular plane can be explained  as a consequence of the spatial anisotropy in the exchange $J$, since the difference between $J_y$ and $J_x$ is of $\sim 0.5$ meV. This can explain why the magnetization does not saturate with a field of several Tesla, since a field of 1 Tesla involves an energy of $g \mu_B B \sim 0.1$ meV for $S=1$.\\

For a field perpendicular to the molecular plane, that is, in the $z$ direction, the Monte Carlo results show that the magnetization is always lower, for any value of $B$, than the average magnetization for $B$ in the molecular plane. Although the results show that $|J_x|<|J_z|<|J_y|$, with $J_x=$1.77 meV, $J_z=$2.07 meV and $J_y=$2.31 meV, which would indicate that it is more favourable for the magnetic moment to move in the {\it{yz}} plane, if positive value of $D=0.65$ meV is taken into account, the final result is that it is energetically more favorable for the magnetic moment to remain in the molecular plane. This explains the result mentioned above.\\

The previous results are in line with magnetization measurements in the $xy$ and $z$ direction \cite{Bartolome2010}, and can also explain why in the non-templated films \cite{Wu2019}, in which the molecules are perpendicular to the substrate, when a field parallel to the substrate is applied, the magnetization is always lower than for the templated films, in which the field is always in the molecular plane. This is because in non-templated films, for some grains, the field is, or almost is, perpendicular to the molecular plane, while for some others it is in the molecular plane, with intermediate situations for other grains, so that the average magnetization can be simulated as an average of the in plane magnetization with the out of plane one.

All in all, our model gives an explanation that can, at least partly, account for the reported results for the non-saturation of the magnetization in FePc thin films. It can be expected that this results also contribute to understand the experiments in FePc powders.

\subsection{Magnetic Solitons}

Ferromagnetic one dimensional systems have magnetic solitons as natural excitations above the ground state \cite{Samalam1982,Wu2019}. 
Solitons possess special properties like stability, mobility without loosing their shape, and energy efficiency, making them promising candidates for various applications in the field of spintronics or information processing. For example, magnetic solitons have been proposed as a mechanism for remotely manipulating single qbits \cite{Cuccoli2016}. Besides, the density of magnetic solitons influences experimental measurements of specific heat \cite{Ramirez82}, neutron scattering experiments \cite{Kjems78}, Mossbauer spectra \cite{Frommen96} and dynamic structure factor \cite{Leung1979}.Several studies in similar model Hamiltonians \cite{Samalam1982, Gaulin1987,Etrich1985} have found this type of excitations. They have also been reported in experimental studies of cuasi 1D compounds, being $Cs Ni F_3$ \cite{Kjems78,Ramirez82} the paradigmatic ferromagnetic soliton bearing compound, and also others like quasi 1D antiferromagnets $Li_2  Mn_{0.98} Fe_{0.02} F_5$ and $Na_2  Mn_{0.98} Fe_{0.02} F_5$ \cite{Frommen96}. In relation to FePc chains, in Ref. \cite{Filoti2006} Mossbauer and a.c. susceptibility measurements, and in Ref. \cite{Wu2019} susceptibility measurements, were interpreted in terms of magnetic solitons, but with different models, or by means of an equation for the energy of the solitons derived from a renormalized Sine-Gordon theory \cite{Samalam1982}. Here, in the context of our first-principles derived model Hamiltonian, we study their existence, type and density, in particular as a function of temperature and magnetic field. \\  

In figure \ref{plotsolitons} the three components of the spins along a segment of 200 sites of our model is plotted, for $T=8 K$ and a magnetic field of $0.25$ Tesla in the 'y' direction. $2 \pi$ solitons can be visualized. The observed solitons are predominantly of the xy type, i.e. the spins remain mostly in the plane of the molecule, but a certain amount of $z$ component is also present. We have also made calculations with $D$ 10 times higher, and checked that the number of solitons changes little but their $z$ component decreases much further. In ferromagnetic isotropic Heisenberg models with single ion anisotropy, solitons are expected to be purely of the xy type for temperatures much lower than $T=(JD)^{1/2}/k_B$, which amounts to between $12$ and $14$ K for the parameters of this model, but $\sim 40$ K for $D$ 10 times higher, and this is the reason why we observe a certain amount of z component in the solitons for the parameters of our model, but almost none for $D$ enhanced by a factor of 10. 

In order to calculate the density of magnetic solitons, the following  quantitative criterion is established, following similar criteria established in previous works \cite{Gaulin1987, Gerling1984}:

\begin{figure}[h]
\includegraphics[width=8.3cm]{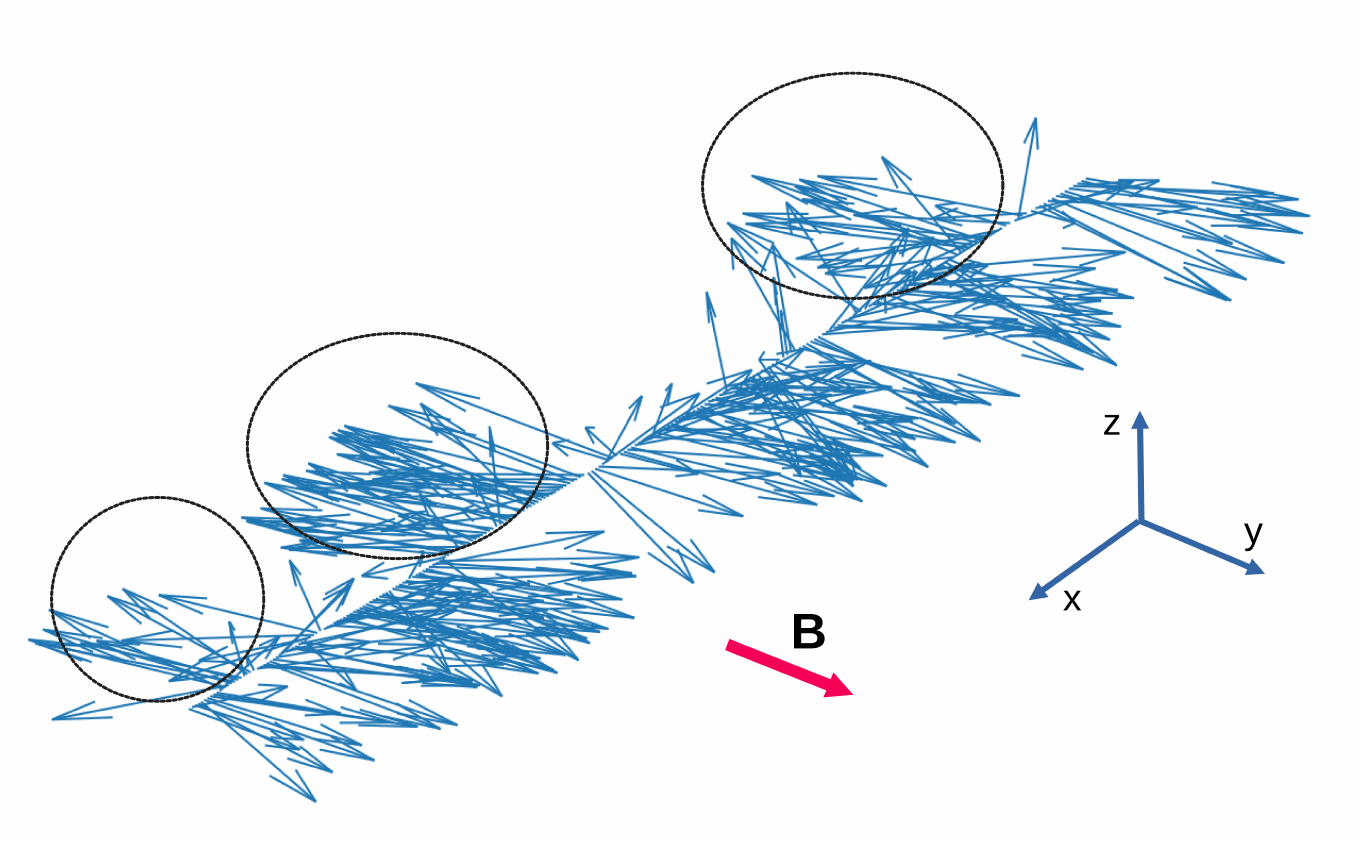}  
\caption{Snapshot of the spins along a selected segment of 200 sites (out of a total of 1000 sites per chain), representing the 1D arrangement of FePC molecules, for $T=8$ K and a magnetic field of $B=0,25$ Tesla. Several magnetic solitons, marked with circles, can be observed.}
\label{plotsolitons}
\end{figure}
 \begin{enumerate}
     \item For a given temperature T we average the y components of all the sites and and obtain $\bar{S_y}$.
     \item A soliton is counted if $\bar{S_y} > 0.4 \times S$ (in our case S=1) and at a certain site $i$ $S_i^y < -0.4 \times \bar{S_y}$, and this remains valid for at least 3 more sites, i.e, the soliton width is larger than 3 sites. Thus we obtain $n_s$.
     \item We then average the $n_s$ values by repeating 2) every $10^6$ MC steps out of a total of $10^9$ steps.
     
 \end{enumerate}

\begin{figure}[h]
\includegraphics[width=8.3cm]{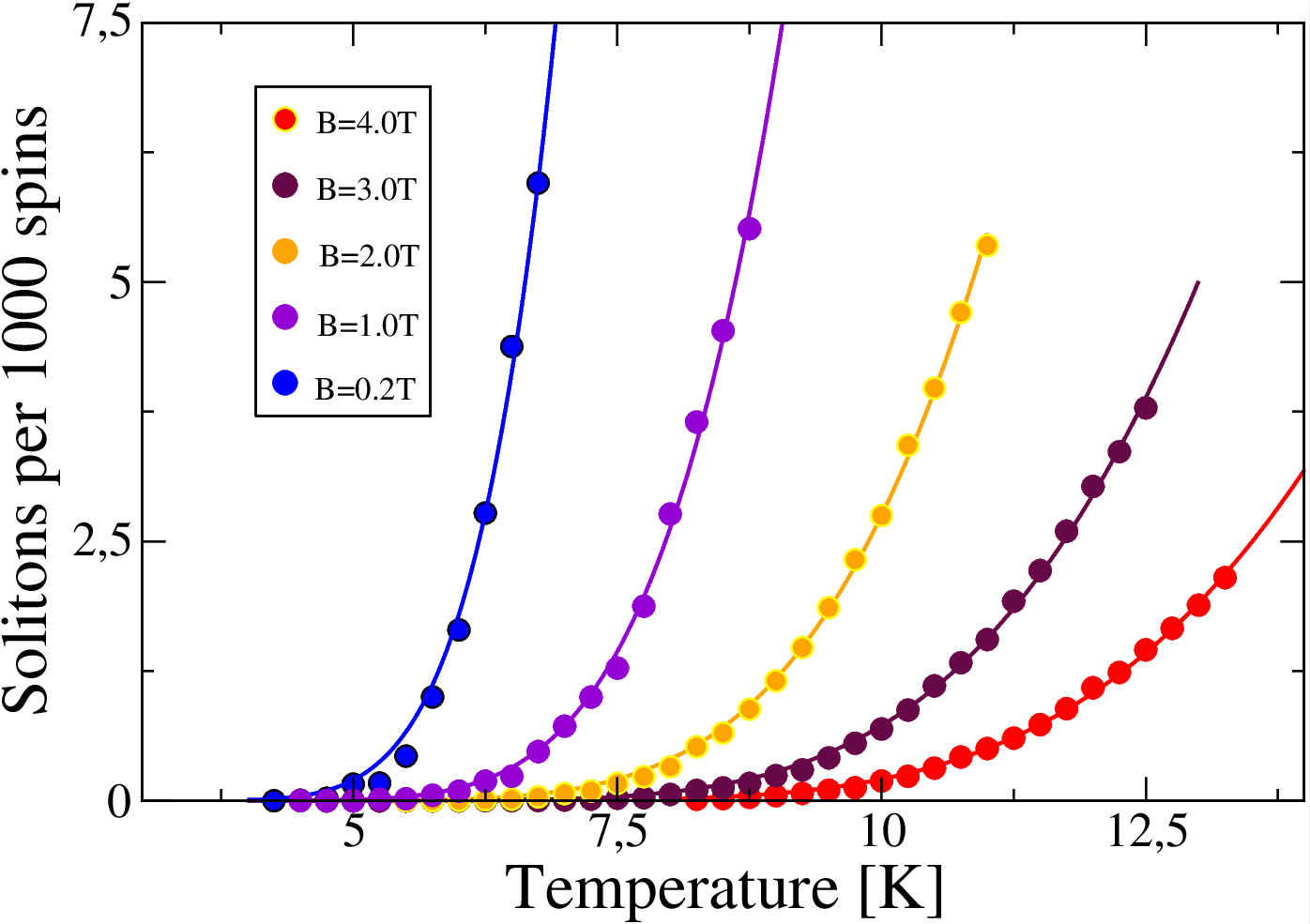}
\caption{Density of magnetic solitons as a function of temperature for several magnetic fields. The lines are fittings with eq. \ref{fitting}. The error bars are covered by the symbols.}  
\label{density_solitons}
\end{figure}

Besides, for low fields, like in Fig. \ref{plotsolitons}, we can roughly count the number of solitons per 100 sites visually, as is sketched in that figure. The criterion above gives the same number as the visual count. With this criterion, the density of magnetic solitons is plotted in Fig. \ref{density_solitons}, together with a fit by an Arrhenius equation \cite{Gaulin1987}. We have tested that across the full range of temperatures and magnetic field strengths investigated, the statistical error associated with our results remained below $1.10^{-2}$ and decreases as $n_s$ decreases.

\begin{equation}
n_s= (A/T) \exp(-E_s/T)
\label{fitting}
\end{equation}

\begin{figure}[h]
 \includegraphics[width=8.3cm]{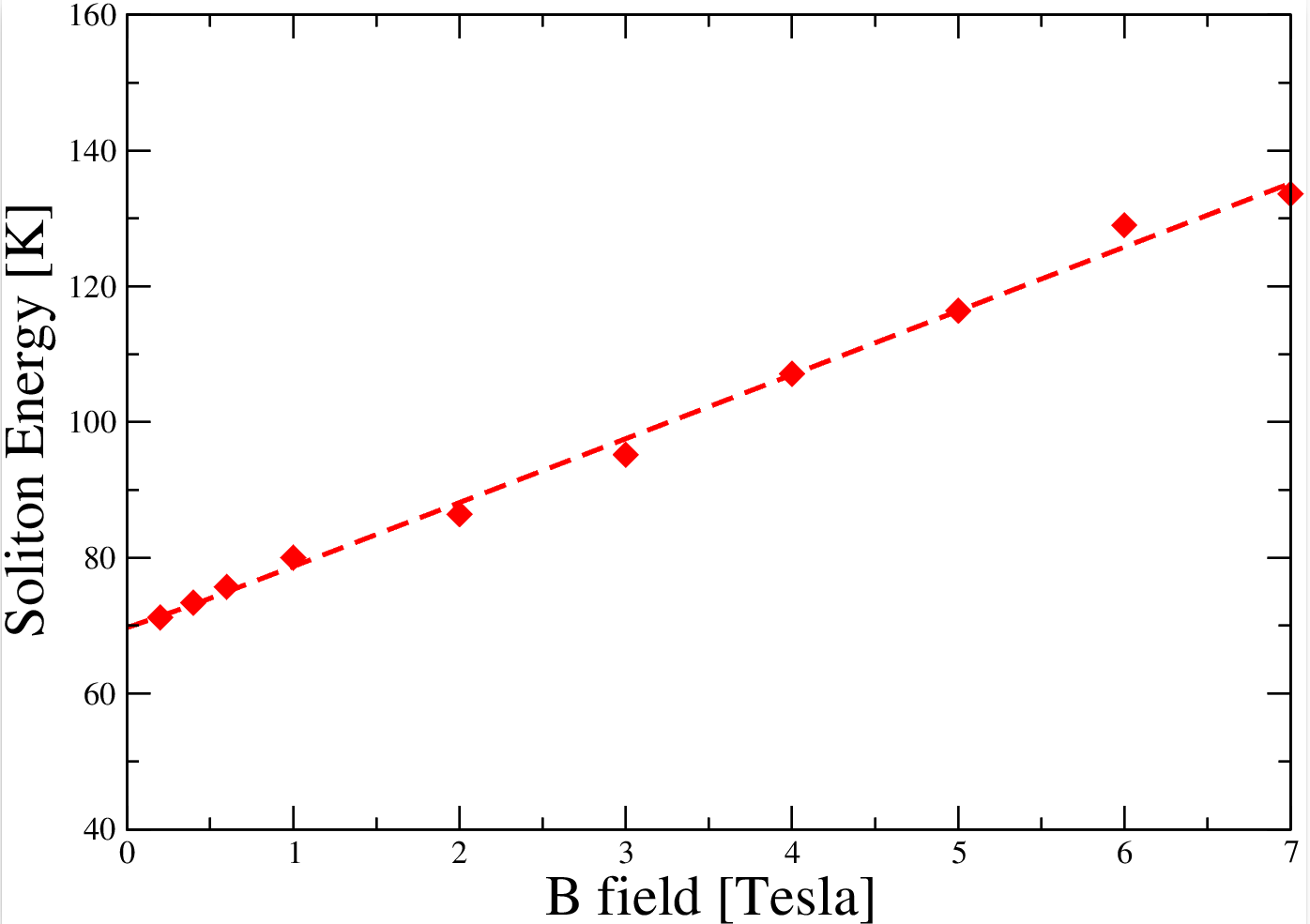}
 \caption{Energy of magnetic solitons as a function of magnetic field according to the fitting done with eq. \ref{fitting}. A linear regression is depicted by a dashed line. } 
 \label{energy_solitons}
 \end{figure}

We can see that the density of solitons increases exponentially as a function of temperature up to the transition temperature. From this temperature on, magnetic solitons decay abruptly to zero (not shown), and this occurs naturally because the magnetic order disappears. The fitting with the Arrhenius equation gives excellent results. Soliton energies are shown in Fig. \ref{energy_solitons} as a function of the magnetic field. For low magnetic fields, the energy is around $73$ K, and grows linearly with the field. This energy is much higher than any of the $J$, and this is in line with what was found in antiferromagnetic Heisenberg models, where it was shown that a small amount of anisotropic exchange leads to the formation of solitons \cite{Gaulin1987}, of energy much higher than $J$. The value obtained for the soliton energy has a very good agreement with what was obtained from Mossbauer/a.c. susceptibility experiments \cite{Filoti2006, Bartolome2015}, i.e. $72$ K with $B=0.08$ Tesla. It is also in between the energy values from the classical domain wall theory $E_s=2 \ \pi \ \sqrt{JD} \simeq 90 K$ and the Sine-Gordon theory $E_s=4 \ \sqrt{JD} \simeq 60 K$ \cite{Frommen96} for an isotropic Heisenberg model with single ion anisotropy $D$, adopting for $J$ the value of $J_y$ for our model. \\

\begin{figure}
 \includegraphics[width=8.3cm]{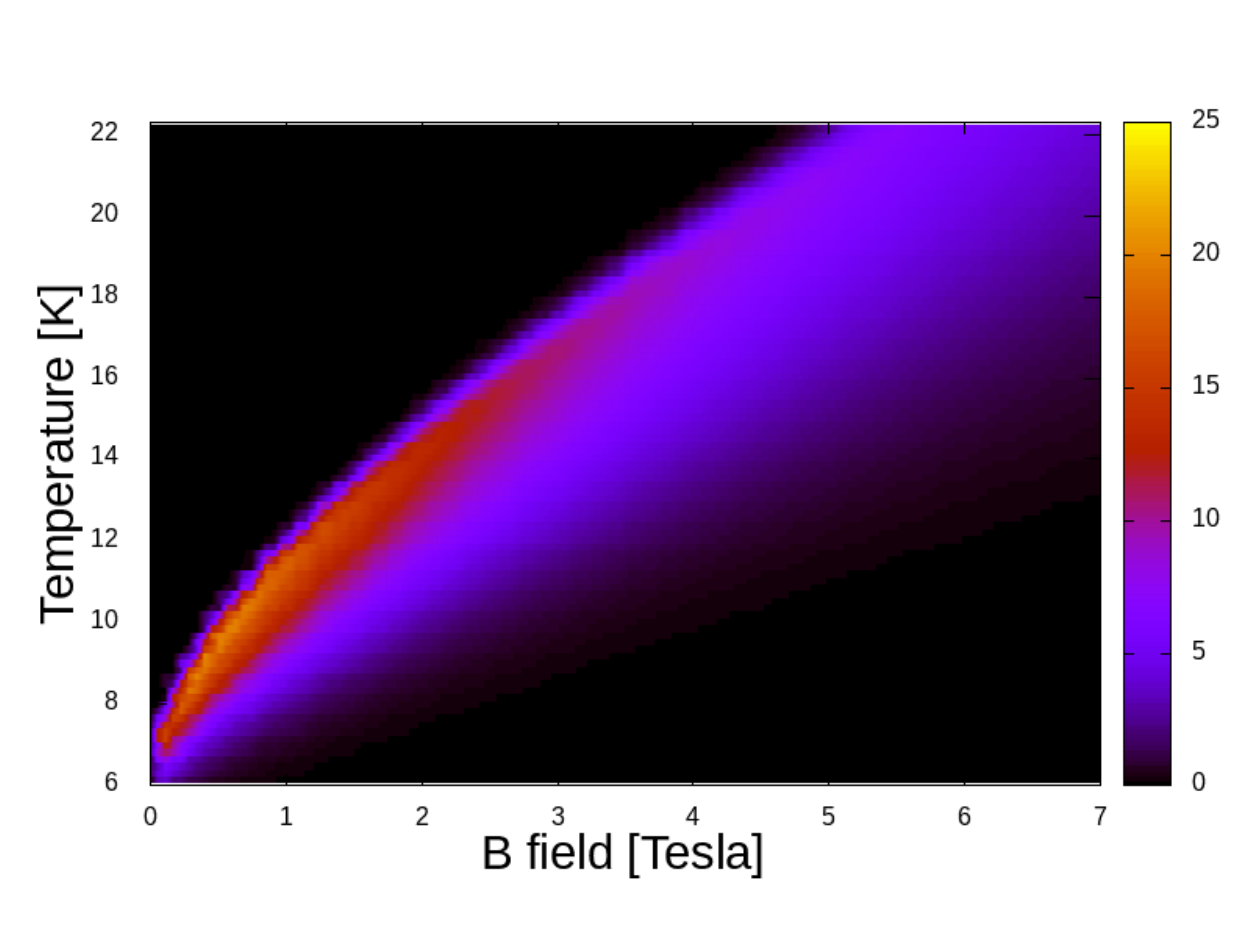}
 \caption{Density of magnetic solitons as a function of temperature for several magnetic fields.} 
 \label{2D_map_dens_solitones}
 \end{figure}

Finally, in Fig. \ref{2D_map_dens_solitones} we plot a colour map for the density of mangetic solitons as a function of magnetic field and temperature. It can be clearly observed how magnetic solitons are excited with temperature and magnetic field up to the transition to the paramagnetic phase. The magnetic field increases the transition temperature (for example at $B=1$ Tesla the transition temperature is $\sim 12 K$ and for $B=4$ Tesla is $\sim 20 K$) and consequently solitons are excited up to higher temperatures as the field increases, but at the same time the soliton density reaches lower values. The highest solitons densities are reached for fields below $2$ Tesla at temperatures between $8$ and $12$ K.  \\ 

All in all, Monte Carlo simulations show the existence of magnetic solitons in the model proposed for $\alpha_+$ FePc in thin films, which are thermally activated in the ferromagnetic phase and increase their energy with the application of a magnetic field, which renders them more difficult to be thermally excitated. \\

\section{Conclusions.}
\label{Conclusions}

In this work we have theoretically studied $\alpha_+$ polymorphism of FePc Phthalocyanines that is found in thin films, consisting of quasi one dimensional chains of FePc molecules. By means of DFT calculations, after reproducing the structural properties of the phase, we have calculated the parameters of a Heisenberg model with single ion anisotropy, finding that due to the spin-orbit interaction the magnetic exchange interaction is anisotropic, i.e. that $J_x \neq J_y \neq J_z$, with $J_x=1.77$ meV, $J_y=2.31$ meV, and $J_z=2.07$ meV. Calculations also show a single ion anisotropy $D$ of $0.65$ meV and negligible $E$.  Adding to this one dimensional model a small interchain hopping, as estimated from experiments and theory \cite{Wu2019,Wu2013}, with classical Monte Carlo calculations we calculated the magnetization at low temperatures, finding that the non saturation found in thin films at fields up to 7 Tesla can be explained by the model. However, in FePc powders, the non-saturation of the magnetization extends to much higher fields, as a consequence of the canted structure of the herringbone arrangement. We have also found that our model is consistent with the existence of magnetic solitons in FePc TF. The solitons observed in MC calculations are of the $2 \pi$ or double kink type. It is found that there is a non-negligible density of solitons even at zero magnetic field, and that the soliton energy for zero magnetic field is around $73$ K, in line with Mossbauer and a.c. susceptibility experiments. Their energy increases linearly with magnetic field, and the thermal energy window in which they are exited is at higher temperatures as the field increases, but at the same time their density also decreases.\\

\section{Conflicts of interest}
There are no conflicts to declare.

\section{Acknowledgements.}
R. P. , P. A and I. H. acknowledge financial support from CONICET PIP0883 and the computational time provided by the CCT-Rosario computational center and Computational Simulation Center (CSC) for Technological Applications, members of the High Performance Computing National System (SNCAD-MincyT Argentina). We acknowledge fruitful conversations with N. O. Vargas, L. O. Manuel, C. J. Gazza and A. E. Trumper.

\bibliography{References}

\providecommand{\latin}[1]{#1}
\makeatletter
\providecommand{\doi}
  {\begingroup\let\do\@makeother\dospecials
  \catcode`\{=1 \catcode`\}=2 \doi@aux}
\providecommand{\doi@aux}[1]{\endgroup\texttt{#1}}
\makeatother
\providecommand*\mcitethebibliography{\thebibliography}
\csname @ifundefined\endcsname{endmcitethebibliography}  {\let\endmcitethebibliography\endthebibliography}{}
\begin{mcitethebibliography}{62}
\providecommand*\natexlab[1]{#1}
\providecommand*\mciteSetBstSublistMode[1]{}
\providecommand*\mciteSetBstMaxWidthForm[2]{}
\providecommand*\mciteBstWouldAddEndPuncttrue
  {\def\EndOfBibitem{\unskip.}}
\providecommand*\mciteBstWouldAddEndPunctfalse
  {\let\EndOfBibitem\relax}
\providecommand*\mciteSetBstMidEndSepPunct[3]{}
\providecommand*\mciteSetBstSublistLabelBeginEnd[3]{}
\providecommand*\EndOfBibitem{}
\mciteSetBstSublistMode{f}
\mciteSetBstMaxWidthForm{subitem}{(\alph{mcitesubitemcount})}
\mciteSetBstSublistLabelBeginEnd
  {\mcitemaxwidthsubitemform\space}
  {\relax}
  {\relax}

\bibitem[Guo \latin{et~al.}(2019)Guo, Gu, Zhu, and Sun]{Guo2019}
Guo,~L.; Gu,~X.; Zhu,~X.; Sun,~X. Recent Advances in Molecular Spintronics: Multifunctional Spintronic Devices. \emph{Advanced Materials} \textbf{2019}, \emph{31}, 1805355\relax
\mciteBstWouldAddEndPuncttrue
\mciteSetBstMidEndSepPunct{\mcitedefaultmidpunct}
{\mcitedefaultendpunct}{\mcitedefaultseppunct}\relax
\EndOfBibitem
\bibitem[Ingan{\"a}s(2018)]{Inganas2018}
Ingan{\"a}s,~O. Organic photovoltaics over three decades. \emph{Advanced materials} \textbf{2018}, \emph{30}, 1800388\relax
\mciteBstWouldAddEndPuncttrue
\mciteSetBstMidEndSepPunct{\mcitedefaultmidpunct}
{\mcitedefaultendpunct}{\mcitedefaultseppunct}\relax
\EndOfBibitem
\bibitem[Grant \latin{et~al.}(2019)Grant, Josey, Sampson, Mudigonda, Bender, and Lessard]{Grant2019}
Grant,~T.~M.; Josey,~D.~S.; Sampson,~K.~L.; Mudigonda,~T.; Bender,~T.~P.; Lessard,~B.~H. Boron Subphthalocyanines and Silicon Phthalocyanines for Use as Active Materials in Organic Photovoltaics. \emph{The Chemical Record} \textbf{2019}, \emph{19}, 1093--1112\relax
\mciteBstWouldAddEndPuncttrue
\mciteSetBstMidEndSepPunct{\mcitedefaultmidpunct}
{\mcitedefaultendpunct}{\mcitedefaultseppunct}\relax
\EndOfBibitem
\bibitem[Fleet \latin{et~al.}(2017)Fleet, Stott, Villis, Din, Serri, Aeppli, Heutz, and Nathan]{Fleet2017}
Fleet,~L.~R.; Stott,~J.; Villis,~B.; Din,~S.; Serri,~M.; Aeppli,~G.; Heutz,~S.; Nathan,~A. Self-Assembled Molecular Nanowires for High-Performance Organic Transistors. \emph{ACS Applied Materials {\&} Interfaces} \textbf{2017}, \emph{9}, 20686--20695\relax
\mciteBstWouldAddEndPuncttrue
\mciteSetBstMidEndSepPunct{\mcitedefaultmidpunct}
{\mcitedefaultendpunct}{\mcitedefaultseppunct}\relax
\EndOfBibitem
\bibitem[Klauk(2010)]{Klauk2010}
Klauk,~H. Organic thin-film transistors. \emph{Chemical Society Reviews} \textbf{2010}, \emph{39}, 2643--2666\relax
\mciteBstWouldAddEndPuncttrue
\mciteSetBstMidEndSepPunct{\mcitedefaultmidpunct}
{\mcitedefaultendpunct}{\mcitedefaultseppunct}\relax
\EndOfBibitem
\bibitem[Zhou \latin{et~al.}(2021)Zhou, Yutronkie, Lessard, and Brusso]{Zhou2021}
Zhou,~W.; Yutronkie,~N.~J.; Lessard,~B.~H.; Brusso,~J.~L. From chemical curiosity to versatile building blocks: unmasking the hidden potential of main-group phthalocyanines in organic field-effect transistors. \emph{Mater. Adv.} \textbf{2021}, \emph{2}, 165--185\relax
\mciteBstWouldAddEndPuncttrue
\mciteSetBstMidEndSepPunct{\mcitedefaultmidpunct}
{\mcitedefaultendpunct}{\mcitedefaultseppunct}\relax
\EndOfBibitem
\bibitem[Kalyani and Dhoble(2012)Kalyani, and Dhoble]{Kalyani2012}
Kalyani,~N.~T.; Dhoble,~S. Organic light emitting diodes: Energy saving lighting technology—A review. \emph{Renewable and Sustainable Energy Reviews} \textbf{2012}, \emph{16}, 2696--2723\relax
\mciteBstWouldAddEndPuncttrue
\mciteSetBstMidEndSepPunct{\mcitedefaultmidpunct}
{\mcitedefaultendpunct}{\mcitedefaultseppunct}\relax
\EndOfBibitem
\bibitem[Plint \latin{et~al.}(2016)Plint, Lessard, and Bender]{Plint2016}
Plint,~T.; Lessard,~B.~H.; Bender,~T.~P. Assessing the potential of group 13 and 14 metal/metalloid phthalocyanines as hole transport layers in organic light emitting diodes. \emph{Journal of Applied Physics} \textbf{2016}, \emph{119}, 145502\relax
\mciteBstWouldAddEndPuncttrue
\mciteSetBstMidEndSepPunct{\mcitedefaultmidpunct}
{\mcitedefaultendpunct}{\mcitedefaultseppunct}\relax
\EndOfBibitem
\bibitem[Rivnay \latin{et~al.}(2018)Rivnay, Inal, Salleo, Owens, Berggren, and Malliaras]{Rivnay2018}
Rivnay,~J.; Inal,~S.; Salleo,~A.; Owens,~R.~M.; Berggren,~M.; Malliaras,~G.~G. Organic electrochemical transistors. \emph{Nature Reviews Materials} \textbf{2018}, \emph{3}, 1--14\relax
\mciteBstWouldAddEndPuncttrue
\mciteSetBstMidEndSepPunct{\mcitedefaultmidpunct}
{\mcitedefaultendpunct}{\mcitedefaultseppunct}\relax
\EndOfBibitem
\bibitem[Basova and Ray(2020)Basova, and Ray]{Basova2020}
Basova,~T.~V.; Ray,~A.~K. Review—Hybrid Materials Based on Phthalocyanines and Metal Nanoparticles for Chemiresistive and Electrochemical Sensors: A Mini-Review. \emph{ECS Journal of Solid State Science and Technology} \textbf{2020}, \emph{9}, 061001\relax
\mciteBstWouldAddEndPuncttrue
\mciteSetBstMidEndSepPunct{\mcitedefaultmidpunct}
{\mcitedefaultendpunct}{\mcitedefaultseppunct}\relax
\EndOfBibitem
\bibitem[Cranston and Lessard(2021)Cranston, and Lessard]{Cranston2021}
Cranston,~R.~R.; Lessard,~B.~H. Metal phthalocyanines: thin-film formation{,} microstructure{,} and physical properties. \emph{RSC Adv.} \textbf{2021}, \emph{11}, 21716--21737\relax
\mciteBstWouldAddEndPuncttrue
\mciteSetBstMidEndSepPunct{\mcitedefaultmidpunct}
{\mcitedefaultendpunct}{\mcitedefaultseppunct}\relax
\EndOfBibitem
\bibitem[Bartolom{\'e} \latin{et~al.}(2014)Bartolom{\'e}, Monton, and Schuller]{Bartolome2014}
Bartolom{\'e},~J.; Monton,~C.; Schuller,~I.~K. In \emph{Molecular Magnets: Physics and Applications}; Bartolom{\'e},~J., Luis,~F., Fern{\'a}ndez,~J.~F., Eds.; Springer Berlin Heidelberg: Berlin, Heidelberg, 2014; pp 221--245\relax
\mciteBstWouldAddEndPuncttrue
\mciteSetBstMidEndSepPunct{\mcitedefaultmidpunct}
{\mcitedefaultendpunct}{\mcitedefaultseppunct}\relax
\EndOfBibitem
\bibitem[Hoshino \latin{et~al.}(2003)Hoshino, Takenaka, and Miyaji]{Hoshiro2003}
Hoshino,~A.; Takenaka,~Y.; Miyaji,~H. Redetermination of the crystal structure of $\alpha$-copper phthalocyanine grown on KCl. \emph{Acta Cryst. B} \textbf{2003}, \emph{59}, 393\relax
\mciteBstWouldAddEndPuncttrue
\mciteSetBstMidEndSepPunct{\mcitedefaultmidpunct}
{\mcitedefaultendpunct}{\mcitedefaultseppunct}\relax
\EndOfBibitem
\bibitem[Wu \latin{et~al.}(2019)Wu, Robaschik, Fleet, Felton, Aeppli, and Heutz]{Wu2019}
Wu,~Z.; Robaschik,~P.; Fleet,~L.~R.; Felton,~S.; Aeppli,~G.; Heutz,~S. Controlling Ferromagnetic Ground States and Solitons in Thin Films and Nanowires Built from Iron Phthalocyanine Chains. \emph{Advanced Functional Materials} \textbf{2019}, \emph{29}, 1902550\relax
\mciteBstWouldAddEndPuncttrue
\mciteSetBstMidEndSepPunct{\mcitedefaultmidpunct}
{\mcitedefaultendpunct}{\mcitedefaultseppunct}\relax
\EndOfBibitem
\bibitem[Yim \latin{et~al.}(2003)Yim, Heutz, and Jones]{Yim2003}
Yim,~S.; Heutz,~S.; Jones,~T.~S. Influence of intermolecular interactions on the structure of phthalocyanine layers in molecular thin film heterostructures. \emph{Phys. Rev. B} \textbf{2003}, \emph{67}, 165308\relax
\mciteBstWouldAddEndPuncttrue
\mciteSetBstMidEndSepPunct{\mcitedefaultmidpunct}
{\mcitedefaultendpunct}{\mcitedefaultseppunct}\relax
\EndOfBibitem
\bibitem[Wang \latin{et~al.}(2010)Wang, Mauthoor, Din, Gardener, Chang, Warner, Aeppli, McComb, Ryan, Wu, Fisher, Stoneham, and Heutz]{Wang2010}
Wang,~H.; Mauthoor,~S.; Din,~S.; Gardener,~J.~A.; Chang,~R.; Warner,~M.; Aeppli,~G.; McComb,~D.~W.; Ryan,~M.~P.; Wu,~W. \latin{et~al.}  Ultralong Copper Phthalocyanine Nanowires with New Crystal Structure and Broad Optical Absorption. \emph{ACS Nano} \textbf{2010}, \emph{4}, 3921--3926, PMID: 20527798\relax
\mciteBstWouldAddEndPuncttrue
\mciteSetBstMidEndSepPunct{\mcitedefaultmidpunct}
{\mcitedefaultendpunct}{\mcitedefaultseppunct}\relax
\EndOfBibitem
\bibitem[Bartolome \latin{et~al.}(2015)Bartolome, Bunau, Garcia, Piantek, Pascual, Schuller, Gredig, Wilhelm, Rogalev, and Bartolome]{Bartolome2015}
Bartolome,~F.; Bunau,~O.; Garcia,~L.~M.; Piantek,~C. R. N.~M.; Pascual,~J.~I.; Schuller,~I.~K.; Gredig,~T.; Wilhelm,~F.; Rogalev,~A.; Bartolome,~J. Molecular tilting and columnar stacking of Fe phthalocyanine thin films on Au(111). \emph{Journal of Applied Physics} \textbf{2015}, \emph{117}, 17A735\relax
\mciteBstWouldAddEndPuncttrue
\mciteSetBstMidEndSepPunct{\mcitedefaultmidpunct}
{\mcitedefaultendpunct}{\mcitedefaultseppunct}\relax
\EndOfBibitem
\bibitem[Ashida \latin{et~al.}(1966)Ashida, Uyeda, and Suito]{Ashida1966}
Ashida,~M.; Uyeda,~N.; Suito,~E. Unit Cell Metastable-form Constants of Various Phthalocyanines. \emph{Bulletin of the Chemical Society of Japan} \textbf{1966}, \emph{39}, 2616--2624\relax
\mciteBstWouldAddEndPuncttrue
\mciteSetBstMidEndSepPunct{\mcitedefaultmidpunct}
{\mcitedefaultendpunct}{\mcitedefaultseppunct}\relax
\EndOfBibitem
\bibitem[Evangelisti \latin{et~al.}(2002)Evangelisti, Bartolom\'e, de~Jongh, and Filoti]{Evangelisti2002}
Evangelisti,~M.; Bartolom\'e,~J.; de~Jongh,~L.~J.; Filoti,~G. Magnetic properties of $\ensuremath{\alpha}$-iron(II) phthalocyanine. \emph{Phys. Rev. B} \textbf{2002}, \emph{66}, 144410\relax
\mciteBstWouldAddEndPuncttrue
\mciteSetBstMidEndSepPunct{\mcitedefaultmidpunct}
{\mcitedefaultendpunct}{\mcitedefaultseppunct}\relax
\EndOfBibitem
\bibitem[Filoti \latin{et~al.}(2006)Filoti, Kuzmin, and Bartolom\'e]{Filoti2006}
Filoti,~G.; Kuzmin,~M.~D.; Bartolom\'e,~J. M\"ossbauer study of the hyperfine interactions and spin dynamics in $\ensuremath{\alpha}$-iron(II) phthalocyanine. \emph{Phys. Rev. B} \textbf{2006}, \emph{74}, 134420\relax
\mciteBstWouldAddEndPuncttrue
\mciteSetBstMidEndSepPunct{\mcitedefaultmidpunct}
{\mcitedefaultendpunct}{\mcitedefaultseppunct}\relax
\EndOfBibitem
\bibitem[Vargas \latin{et~al.}(2020)Vargas, Torres, Baker, Lee, Kiwi, Willey, Monton, and Schuller]{Vargas2020}
Vargas,~N.~M.; Torres,~F.; Baker,~A.~A.; Lee,~J. R.~I.; Kiwi,~M.; Willey,~T.~M.; Monton,~C.; Schuller,~I.~K. {Helical spin structure in iron chains with hybridized boundaries}. \emph{Applied Physics Letters} \textbf{2020}, \emph{117}, 213105\relax
\mciteBstWouldAddEndPuncttrue
\mciteSetBstMidEndSepPunct{\mcitedefaultmidpunct}
{\mcitedefaultendpunct}{\mcitedefaultseppunct}\relax
\EndOfBibitem
\bibitem[Gredig \latin{et~al.}(2012)Gredig, Werber, Guerra, Silverstein, Byrne, and Cacha]{Gredig2012}
Gredig,~T.; Werber,~M.; Guerra,~J.~L.; Silverstein,~E.~A.; Byrne,~M.~P.; Cacha,~B.~G. Coercivity Control of Variable-Length Iron Chains in Phthalocyanine Thin Films. \emph{Journal of Superconductivity and Novel Magnetism} \textbf{2012}, \emph{25}, 2199--2203\relax
\mciteBstWouldAddEndPuncttrue
\mciteSetBstMidEndSepPunct{\mcitedefaultmidpunct}
{\mcitedefaultendpunct}{\mcitedefaultseppunct}\relax
\EndOfBibitem
\bibitem[Kresse and Hafner(1993)Kresse, and Hafner]{Kresse1993a}
Kresse,~G.; Hafner,~J. Ab Initio Molecular Dynamics for Liquid Metals. \emph{Phys. Rev. B} \textbf{1993}, \emph{47}, 558\relax
\mciteBstWouldAddEndPuncttrue
\mciteSetBstMidEndSepPunct{\mcitedefaultmidpunct}
{\mcitedefaultendpunct}{\mcitedefaultseppunct}\relax
\EndOfBibitem
\bibitem[Kresse and Hafner(1993)Kresse, and Hafner]{Kresse1993b}
Kresse,~G.; Hafner,~J. Ab Initio Molecular Dynamics for Open-Shell Transition Metals. \emph{Phys. Rev. B} \textbf{1993}, \emph{48}, 13115\relax
\mciteBstWouldAddEndPuncttrue
\mciteSetBstMidEndSepPunct{\mcitedefaultmidpunct}
{\mcitedefaultendpunct}{\mcitedefaultseppunct}\relax
\EndOfBibitem
\bibitem[Kresse and Furthm\"{u}ller(1996)Kresse, and Furthm\"{u}ller]{Kresse1996a}
Kresse,~G.; Furthm\"{u}ller,~J. Efficiency of Ab-Initio Total Energy Calculations for Metals and Semiconductors Using a Plane-Wave Basis Set. \emph{Comput. Mater. Sci.} \textbf{1996}, \emph{6}, 15\relax
\mciteBstWouldAddEndPuncttrue
\mciteSetBstMidEndSepPunct{\mcitedefaultmidpunct}
{\mcitedefaultendpunct}{\mcitedefaultseppunct}\relax
\EndOfBibitem
\bibitem[Kresse and Furthm\"{u}ller(1996)Kresse, and Furthm\"{u}ller]{Kresse1996b}
Kresse,~G.; Furthm\"{u}ller,~J. Efficient Iterative Schemes for Ab Initio Total-Energy Calculations Using a Plane-Wave Basis Set. \emph{Phys. Rev. B} \textbf{1996}, \emph{54}, 11169\relax
\mciteBstWouldAddEndPuncttrue
\mciteSetBstMidEndSepPunct{\mcitedefaultmidpunct}
{\mcitedefaultendpunct}{\mcitedefaultseppunct}\relax
\EndOfBibitem
\bibitem[Kresse and Joubert(1999)Kresse, and Joubert]{Kresse1999}
Kresse,~G.; Joubert,~D. From Ultrasoft Pseudopotentials to the Projector Augmented-Wave Method. \emph{Phys. Rev. B} \textbf{1999}, \emph{59}, 1758\relax
\mciteBstWouldAddEndPuncttrue
\mciteSetBstMidEndSepPunct{\mcitedefaultmidpunct}
{\mcitedefaultendpunct}{\mcitedefaultseppunct}\relax
\EndOfBibitem
\bibitem[Hafner(2008)]{Hafner2008}
Hafner,~J. Ab-Initio Simulations of Materials Using VASP: Density-Functional Theory and Beyond. \emph{J.\ Comput.\ Chem.} \textbf{2008}, \emph{29}, 2044--2078\relax
\mciteBstWouldAddEndPuncttrue
\mciteSetBstMidEndSepPunct{\mcitedefaultmidpunct}
{\mcitedefaultendpunct}{\mcitedefaultseppunct}\relax
\EndOfBibitem
\bibitem[Grimme \latin{et~al.}(2010)Grimme, Ehrlich, , and Krieg]{Grimme2010}
Grimme,~S.; Ehrlich,~J. A.~S.; ; Krieg,~H. A consistent and accurate ab initio parametrization of density functional dispersion correction (DFT-D) for the 94 elements H-Pu. \emph{J. Chem. Phys.} \textbf{2010}, \emph{132}, 154104\relax
\mciteBstWouldAddEndPuncttrue
\mciteSetBstMidEndSepPunct{\mcitedefaultmidpunct}
{\mcitedefaultendpunct}{\mcitedefaultseppunct}\relax
\EndOfBibitem
\bibitem[J.Klimes \latin{et~al.}(2010)J.Klimes, Bowler, and Michaelides]{Klimes2011}
J.Klimes; Bowler,~D.~R.; Michaelides,~A. Van der Waals density functionals applied to solids. \emph{Phys. Rev. B} \textbf{2010}, \emph{83}, 195133\relax
\mciteBstWouldAddEndPuncttrue
\mciteSetBstMidEndSepPunct{\mcitedefaultmidpunct}
{\mcitedefaultendpunct}{\mcitedefaultseppunct}\relax
\EndOfBibitem
\bibitem[Dudarev \latin{et~al.}(1998)Dudarev, Botton, Savrasov, Humphreys, and Sutton]{Dudarev1998}
Dudarev,~S.~L.; Botton,~G.~A.; Savrasov,~S.~Y.; Humphreys,~C.~J.; Sutton,~A.~P. Electron energy loss. \emph{Phys. Rev. B} \textbf{1998}, \emph{57}, 1505\relax
\mciteBstWouldAddEndPuncttrue
\mciteSetBstMidEndSepPunct{\mcitedefaultmidpunct}
{\mcitedefaultendpunct}{\mcitedefaultseppunct}\relax
\EndOfBibitem
\bibitem[Mugarza \latin{et~al.}(2012)Mugarza, Robles, Krull, Koryt\'ar, Lorente, and Gambardella]{Roberto2012}
Mugarza,~A.; Robles,~R.; Krull,~C.; Koryt\'ar,~R.; Lorente,~N.; Gambardella,~P. Electronic and magnetic properties of molecule-metal interfaces: Transition-metal phthalocyanines adsorbed on Ag(100). \emph{Phys. Rev. B} \textbf{2012}, \emph{85}, 155437\relax
\mciteBstWouldAddEndPuncttrue
\mciteSetBstMidEndSepPunct{\mcitedefaultmidpunct}
{\mcitedefaultendpunct}{\mcitedefaultseppunct}\relax
\EndOfBibitem
\bibitem[Wang \latin{et~al.}(2018)Wang, Li, Zheng, and Yang]{Wang2018}
Wang,~Y.; Li,~X.; Zheng,~X.; Yang,~J. Manipulation of spin and magnetic anisotropy in bilayer magnetic molecular junctions. \emph{Phys. Chem. Chem. Phys.} \textbf{2018}, \emph{20}, 26396--26404\relax
\mciteBstWouldAddEndPuncttrue
\mciteSetBstMidEndSepPunct{\mcitedefaultmidpunct}
{\mcitedefaultendpunct}{\mcitedefaultseppunct}\relax
\EndOfBibitem
\bibitem[Marom and Kronik(2009)Marom, and Kronik]{10Marom2009}
Marom,~N.; Kronik,~L. Density functional theory of transition metal phthalocyanines, II: electronic structure of MnPc and FePc---symmetry and symmetry breaking. \emph{Applied Physics A} \textbf{2009}, \emph{95}, 165--172\relax
\mciteBstWouldAddEndPuncttrue
\mciteSetBstMidEndSepPunct{\mcitedefaultmidpunct}
{\mcitedefaultendpunct}{\mcitedefaultseppunct}\relax
\EndOfBibitem
\bibitem[Białek \latin{et~al.}(2003)Białek, Kim, and Lee]{40Bialek2003}
Białek,~B.; Kim,~I.~G.; Lee,~J.~I. First-principles study on the electronic structures of iron phthalocyanine monolayer. \emph{Surface Science} \textbf{2003}, \emph{526}, 367--374\relax
\mciteBstWouldAddEndPuncttrue
\mciteSetBstMidEndSepPunct{\mcitedefaultmidpunct}
{\mcitedefaultendpunct}{\mcitedefaultseppunct}\relax
\EndOfBibitem
\bibitem[Kuz'min \latin{et~al.}(2009)Kuz'min, Hayn, and Oison]{41Kuzmin2009}
Kuz'min,~M.~D.; Hayn,~R.; Oison,~V. Ab initio calculated XANES and XMCD spectra of Fe(II) phthalocyanine. \emph{Phys. Rev. B} \textbf{2009}, \emph{79}, 024413\relax
\mciteBstWouldAddEndPuncttrue
\mciteSetBstMidEndSepPunct{\mcitedefaultmidpunct}
{\mcitedefaultendpunct}{\mcitedefaultseppunct}\relax
\EndOfBibitem
\bibitem[Brena \latin{et~al.}(2011)Brena, Puglia, de~Simone, Coreno, Tarafder, Feyer, Banerjee, Göthelid, Sanyal, Oppeneer, and Eriksson]{Brena2011}
Brena,~B.; Puglia,~C.; de~Simone,~M.; Coreno,~M.; Tarafder,~K.; Feyer,~V.; Banerjee,~R.; Göthelid,~E.; Sanyal,~B.; Oppeneer,~P.~M. \latin{et~al.}  {Valence-band electronic structure of iron phthalocyanine: An experimental and theoretical photoelectron spectroscopy study}. \emph{The Journal of Chemical Physics} \textbf{2011}, \emph{134}, 074312\relax
\mciteBstWouldAddEndPuncttrue
\mciteSetBstMidEndSepPunct{\mcitedefaultmidpunct}
{\mcitedefaultendpunct}{\mcitedefaultseppunct}\relax
\EndOfBibitem
\bibitem[Marom and Kronik(2009)Marom, and Kronik]{Marom2009}
Marom,~N.; Kronik,~L. L. Density functional theory of transition metal phthalocyanines, II: electronic structure of MnPc and FePc- symmetry and symmetry breaking. \emph{Appl. Phys.} \textbf{2009}, \emph{95}, 165--172\relax
\mciteBstWouldAddEndPuncttrue
\mciteSetBstMidEndSepPunct{\mcitedefaultmidpunct}
{\mcitedefaultendpunct}{\mcitedefaultseppunct}\relax
\EndOfBibitem
\bibitem[Ichibha \latin{et~al.}(2017)Ichibha, Hou, Hongo, and Maezono1]{Ichibha2017}
Ichibha,~T.; Hou,~Z.; Hongo,~K.; Maezono1,~R. New Insight into the Ground State of FePc: A Diffusion Monte Carlo Study. \emph{Sci. Rep.} \textbf{2017}, \emph{7}, 2011\relax
\mciteBstWouldAddEndPuncttrue
\mciteSetBstMidEndSepPunct{\mcitedefaultmidpunct}
{\mcitedefaultendpunct}{\mcitedefaultseppunct}\relax
\EndOfBibitem
\bibitem[Phung \latin{et~al.}(2023)Phung, Nam, and Saitow]{Phung2023}
Phung,~Q.~M.; Nam,~H.~N.; Saitow,~M. Unraveling the Spin-State Energetics of FeN$_4$ Complexes with Ab Initio Methods. \emph{The Journal of Physical Chemistry A} \textbf{2023}, \emph{127}, 7544--7556\relax
\mciteBstWouldAddEndPuncttrue
\mciteSetBstMidEndSepPunct{\mcitedefaultmidpunct}
{\mcitedefaultendpunct}{\mcitedefaultseppunct}\relax
\EndOfBibitem
\bibitem[Nakamura \latin{et~al.}(2012)Nakamura, Kitaoka, Akiyama, and Ito]{Nakamura2012}
Nakamura,~K.; Kitaoka,~Y.; Akiyama,~T.; Ito,~T. Constraint density functional calculations for multiplets in a ligand-field applied to Fe-phthalocyanine. \emph{Phys. Rev. B} \textbf{2012}, \emph{85}, 235129\relax
\mciteBstWouldAddEndPuncttrue
\mciteSetBstMidEndSepPunct{\mcitedefaultmidpunct}
{\mcitedefaultendpunct}{\mcitedefaultseppunct}\relax
\EndOfBibitem
\bibitem[Xia \latin{et~al.}(2023)Xia, Li, Fang, Jones, and Yang]{Xia2023}
Xia,~H.; Li,~L.; Fang,~M.; Jones,~T.~S.; Yang,~J. Molecular-orientation-dependent magnetic properties of iron phthalocyanine (FePc) thin films and microwires. \emph{Organic Electronics} \textbf{2023}, \emph{121}, 106870\relax
\mciteBstWouldAddEndPuncttrue
\mciteSetBstMidEndSepPunct{\mcitedefaultmidpunct}
{\mcitedefaultendpunct}{\mcitedefaultseppunct}\relax
\EndOfBibitem
\bibitem[Bartolom\'e \latin{et~al.}(2010)Bartolom\'e, Bartolom\'e, Garc\'{\i}a, Filoti, Gredig, Colesniuc, Schuller, and Cezar]{Bartolome2010}
Bartolom\'e,~J.; Bartolom\'e,~F.; Garc\'{\i}a,~L.~M.; Filoti,~G.; Gredig,~T.; Colesniuc,~C.~N.; Schuller,~I.~K.; Cezar,~J.~C. Highly unquenched orbital moment in textured Fe-phthalocyanine thin films. \emph{Phys. Rev. B} \textbf{2010}, \emph{81}, 195405\relax
\mciteBstWouldAddEndPuncttrue
\mciteSetBstMidEndSepPunct{\mcitedefaultmidpunct}
{\mcitedefaultendpunct}{\mcitedefaultseppunct}\relax
\EndOfBibitem
\bibitem[Natoli \latin{et~al.}(2018)Natoli, Kr\"uger, Bartolom\'e, and Bartolom\'e]{Natoli2018}
Natoli,~C.~R.; Kr\"uger,~P.; Bartolom\'e,~J.; Bartolom\'e,~F. Determination of the ground state of an Au-supported FePc film based on the interpretation of Fe $K$- and $L$-edge x-ray magnetic circular dichroism measurements. \emph{Phys. Rev. B} \textbf{2018}, \emph{97}, 155139\relax
\mciteBstWouldAddEndPuncttrue
\mciteSetBstMidEndSepPunct{\mcitedefaultmidpunct}
{\mcitedefaultendpunct}{\mcitedefaultseppunct}\relax
\EndOfBibitem
\bibitem[Barraclough \latin{et~al.}(1970)Barraclough, Martin, Mitra, and Sherwood]{Barraclough1970}
Barraclough,~C.~G.; Martin,~R.~L.; Mitra,~S.; Sherwood,~R.~C. Paramagnetic Anisotropy, Low Temperature Magnetization, and Electronic Structure of Iron(II) Phthalocyanine. \emph{The Journal of Chemical Physics} \textbf{1970}, \emph{53}, 1643--1648\relax
\mciteBstWouldAddEndPuncttrue
\mciteSetBstMidEndSepPunct{\mcitedefaultmidpunct}
{\mcitedefaultendpunct}{\mcitedefaultseppunct}\relax
\EndOfBibitem
\bibitem[Torres \latin{et~al.}(2020)Torres, Kiwi, Vargas, Monton, and Schuller]{Vargas_2020_2}
Torres,~F.; Kiwi,~M.; Vargas,~N.~M.; Monton,~C.; Schuller,~I.~K. Chiral symmetry and scale invariance breaking in spin chains. \emph{AIP Advances} \textbf{2020}, \emph{10}, 025215\relax
\mciteBstWouldAddEndPuncttrue
\mciteSetBstMidEndSepPunct{\mcitedefaultmidpunct}
{\mcitedefaultendpunct}{\mcitedefaultseppunct}\relax
\EndOfBibitem
\bibitem[Wu \latin{et~al.}(2013)Wu, Harrison, and Fisher]{Wu2013}
Wu,~W.; Harrison,~N.~M.; Fisher,~A.~J. Suitability of chromium phthalocyanines to test Haldane's conjecture: First-principles calculations. \emph{Phys. Rev. B} \textbf{2013}, \emph{88}, 224417\relax
\mciteBstWouldAddEndPuncttrue
\mciteSetBstMidEndSepPunct{\mcitedefaultmidpunct}
{\mcitedefaultendpunct}{\mcitedefaultseppunct}\relax
\EndOfBibitem
\bibitem[Moriya(1960)]{Moriya60}
Moriya,~T. Anisotropic Superexchange Interaction and Weak Ferromagnetism. \emph{Phys. Rev.} \textbf{1960}, \emph{120}, 91--98\relax
\mciteBstWouldAddEndPuncttrue
\mciteSetBstMidEndSepPunct{\mcitedefaultmidpunct}
{\mcitedefaultendpunct}{\mcitedefaultseppunct}\relax
\EndOfBibitem
\bibitem[Guterding \latin{et~al.}(2016)Guterding, Valent\'{\i}, and Jeschke]{Guterding2016}
Guterding,~D.; Valent\'{\i},~R.; Jeschke,~H.~O. Reduction of magnetic interlayer coupling in barlowite through isoelectronic substitution. \emph{Phys. Rev. B} \textbf{2016}, \emph{94}, 125136\relax
\mciteBstWouldAddEndPuncttrue
\mciteSetBstMidEndSepPunct{\mcitedefaultmidpunct}
{\mcitedefaultendpunct}{\mcitedefaultseppunct}\relax
\EndOfBibitem
\bibitem[Whangbo and Xiang(2017)Whangbo, and Xiang]{Whangbo2017}
Whangbo,~M.-H.; Xiang,~H. \emph{Handbook of Solid State Chemistry}; John Wiley and Sons, Ltd, 2017; Chapter 10, pp 285--343\relax
\mciteBstWouldAddEndPuncttrue
\mciteSetBstMidEndSepPunct{\mcitedefaultmidpunct}
{\mcitedefaultendpunct}{\mcitedefaultseppunct}\relax
\EndOfBibitem
\bibitem[W{\"a}ckerlin \latin{et~al.}(2022)W{\"a}ckerlin, Cahl{\'i}k, Goikoetxea, Stetsovych, Medvedeva, Redondo, {\v{S}}vec, Delley, Ondr{\'a}{\v{c}}ek, Pinar, Blanco-Rey, Koloren{\v{c}}, Arnau, and Jel{\'i}nek]{Wackerlin2022}
W{\"a}ckerlin,~C.; Cahl{\'i}k,~A.; Goikoetxea,~J.; Stetsovych,~O.; Medvedeva,~D.; Redondo,~J.; {\v{S}}vec,~M.; Delley,~B.; Ondr{\'a}{\v{c}}ek,~M.; Pinar,~A. \latin{et~al.}  Role of the Magnetic Anisotropy in Atomic-Spin Sensing of 1D Molecular Chains. \emph{ACS Nano} \textbf{2022}, \emph{16}, 16402--16413\relax
\mciteBstWouldAddEndPuncttrue
\mciteSetBstMidEndSepPunct{\mcitedefaultmidpunct}
{\mcitedefaultendpunct}{\mcitedefaultseppunct}\relax
\EndOfBibitem
\bibitem[Gallardo \latin{et~al.}(2019)Gallardo, Arnau, Delgado, Baltic, Singha, Donati, Wäckerlin, Dreiser, Rusponi, and Brune]{Gallardo_2019}
Gallardo,~I.; Arnau,~A.; Delgado,~F.; Baltic,~R.; Singha,~A.; Donati,~F.; Wäckerlin,~C.; Dreiser,~J.; Rusponi,~S.; Brune,~H. Large effect of metal substrate on magnetic anisotropy of Co on hexagonal boron nitride. \emph{New Journal of Physics} \textbf{2019}, \emph{21}, 073053\relax
\mciteBstWouldAddEndPuncttrue
\mciteSetBstMidEndSepPunct{\mcitedefaultmidpunct}
{\mcitedefaultendpunct}{\mcitedefaultseppunct}\relax
\EndOfBibitem
\bibitem[Samalam and Kumar(1982)Samalam, and Kumar]{Samalam1982}
Samalam,~V.~K.; Kumar,~P. Statistical mechanics of a magnetic chain. \emph{Phys. Rev. B} \textbf{1982}, \emph{26}, 5146--5152\relax
\mciteBstWouldAddEndPuncttrue
\mciteSetBstMidEndSepPunct{\mcitedefaultmidpunct}
{\mcitedefaultendpunct}{\mcitedefaultseppunct}\relax
\EndOfBibitem
\bibitem[Cuccoli \latin{et~al.}(2016)Cuccoli, Nuzzi, Vaia, and Verrucchi]{Cuccoli2016}
Cuccoli,~A.; Nuzzi,~D.; Vaia,~R.; Verrucchi,~P. Single-qubit remote manipulation by magnetic solitons. \emph{Journal of Magnetism and Magnetic Materials} \textbf{2016}, \emph{400}, 149--153, Proceedings of the 20th International Conference on Magnetism (Barcelona) 5-10 July 2015\relax
\mciteBstWouldAddEndPuncttrue
\mciteSetBstMidEndSepPunct{\mcitedefaultmidpunct}
{\mcitedefaultendpunct}{\mcitedefaultseppunct}\relax
\EndOfBibitem
\bibitem[Ramirez and Wolf(1982)Ramirez, and Wolf]{Ramirez82}
Ramirez,~A.~P.; Wolf,~W.~P. Specific Heat of ${\mathrm{CsNiF}}_{3}$: Evidence for Spin Solitons. \emph{Phys. Rev. Lett.} \textbf{1982}, \emph{49}, 227--230\relax
\mciteBstWouldAddEndPuncttrue
\mciteSetBstMidEndSepPunct{\mcitedefaultmidpunct}
{\mcitedefaultendpunct}{\mcitedefaultseppunct}\relax
\EndOfBibitem
\bibitem[Kjems and Steiner(1978)Kjems, and Steiner]{Kjems78}
Kjems,~J.~K.; Steiner,~M. Evidence for Soliton Modes in the One-Dimensional Ferromagnet CsNi${\mathrm{F}}_{3}$. \emph{Phys. Rev. Lett.} \textbf{1978}, \emph{41}, 1137--1140\relax
\mciteBstWouldAddEndPuncttrue
\mciteSetBstMidEndSepPunct{\mcitedefaultmidpunct}
{\mcitedefaultendpunct}{\mcitedefaultseppunct}\relax
\EndOfBibitem
\bibitem[Frommen \latin{et~al.}(1996)Frommen, Mangold, and Pebler]{Frommen96}
Frommen,~C.; Mangold,~M.; Pebler,~J. Magnetic Solitons in the 1-D Antiferromagnetic Chains of Li2Mn0.98Fe0.02F5 and Na2Mn0.98Fe0.02F5. \emph{Zeitschrift für Naturforschung A} \textbf{1996}, \emph{51}, 939--949\relax
\mciteBstWouldAddEndPuncttrue
\mciteSetBstMidEndSepPunct{\mcitedefaultmidpunct}
{\mcitedefaultendpunct}{\mcitedefaultseppunct}\relax
\EndOfBibitem
\bibitem[Leung and Huber(1979)Leung, and Huber]{Leung1979}
Leung,~K.; Huber,~D. Soliton dynamic structure factors in a planar ferromagnetic chain. \emph{Solid State Communications} \textbf{1979}, \emph{32}, 127--130\relax
\mciteBstWouldAddEndPuncttrue
\mciteSetBstMidEndSepPunct{\mcitedefaultmidpunct}
{\mcitedefaultendpunct}{\mcitedefaultseppunct}\relax
\EndOfBibitem
\bibitem[Gaulin(1987)]{Gaulin1987}
Gaulin,~B.~D. Soliton spin configurations along the classical anisotropic Heisenberg chain. \emph{Journal of Applied Physics} \textbf{1987}, \emph{61}, 4435--4437\relax
\mciteBstWouldAddEndPuncttrue
\mciteSetBstMidEndSepPunct{\mcitedefaultmidpunct}
{\mcitedefaultendpunct}{\mcitedefaultseppunct}\relax
\EndOfBibitem
\bibitem[Etrich \latin{et~al.}(1985)Etrich, Mikeska, Magyari, Thomas, and Weber]{Etrich1985}
Etrich,~C.; Mikeska,~H.~J.; Magyari,~E.; Thomas,~H.; Weber,~R. Solitons on a discrete ferromagnetic spin chain. \emph{Zeitschrift f{\"u}r Physik B Condensed Matter} \textbf{1985}, \emph{62}, 97--111\relax
\mciteBstWouldAddEndPuncttrue
\mciteSetBstMidEndSepPunct{\mcitedefaultmidpunct}
{\mcitedefaultendpunct}{\mcitedefaultseppunct}\relax
\EndOfBibitem
\bibitem[Gerling and Landau(1984)Gerling, and Landau]{Gerling1984}
Gerling,~R.; Landau,~D. Spin solitons in the classical xy-chain. \emph{Journal of Magnetism and Magnetic Materials} \textbf{1984}, \emph{45}, 267--271\relax
\mciteBstWouldAddEndPuncttrue
\mciteSetBstMidEndSepPunct{\mcitedefaultmidpunct}
{\mcitedefaultendpunct}{\mcitedefaultseppunct}\relax
\EndOfBibitem
\end{mcitethebibliography}

\end{document}